\documentclass[12pt,preprint]{aastex}
\usepackage{natbib, epsfig}
\def\beq{\begin{equation}}
\def\enq{\end{equation}}
\def\bea{\begin{array}}
\def\ena{\end{array}}

\newcommand{\siml}{\lower4pt \hbox{$\buildrel < \over \sim$}}
\newcommand{\simg}{\lower4pt \hbox{$\buildrel > \over \sim$}}
\newcommand{\xray}{\mbox{X-ray}}

\newcommand{\g}{\gamma}
\newcommand{\kev}{keV}
\newcommand{\uJy}{\mbox{$\mu$Jy}}
\newcommand{\percmcube}{\mbox{cm$^{-3}$}}
\newcommand{\swift}{{\it Swift}}

\newcommand{\hstlong}{\textit{Hubble Space Telescope}}
\newcommand{\spitzerlong}{\textit{Spitzer Space Telescope}}
\bibpunct[,]{(}{)}{;}{a}{}{,}

\begin{document}

\title{Modeling GRB\,050904: \\
       Autopsy of a Massive Stellar Explosion at $z=6.29$}

\author{ L.~J. Gou\altaffilmark{1}, D.~B. Fox\altaffilmark{1}, and P.
M\'{e}sz\'{a}ros\altaffilmark{1,2}}

\altaffiltext{1}{Department of Astronomy \& Astrophysics, 525 Davey
Laboratory, Pennsylvania State University, University Park, PA
16802}

\altaffiltext{2}{Department of Physics, Pennsylvania State
University, University Park, PA 16802}

\begin{abstract}

GRB\,050904 at redshift $z=6.29$, discovered and observed by \swift\
and with spectroscopic redshift from the Subaru telescope, is the
first gamma-ray burst to be identified from beyond the epoch of
reionization.  Since the progenitors of long gamma-ray bursts have
been identified as massive stars, this event offers a unique
opportunity to investigate star formation environments at this
epoch. Apart from its record redshift, the burst is remarkable in
two respects: first, it exhibits fast-evolving \xray\ and optical
flares that peak simultaneously at $t\approx 470$\,s in the observer
frame, and may thus originate in the same emission region; and
second, its afterglow exhibits an accelerated decay in the
near-infrared (NIR) from $t\approx 10^4$\,s to $t\approx 3\times
10^4$\,s after the burst, coincident with repeated and energetic
\xray\ flaring activity.  We make a complete analysis of available
\xray, NIR, and radio observations, utilizing afterglow models that
incorporate a range of physical effects not previously considered
for this or any other GRB afterglow, and quantifying our model
uncertainties in detail via Markov Chain Monte Carlo analysis.  In
the process, we explore the possibility that the early optical and
\xray\ flare is due to synchrotron and inverse Compton emission from
the reverse shock regions of the outflow.  We suggest that the
period of accelerated decay in the NIR may be due to suppression of
synchrotron radiation by inverse Compton interaction of \xray\ flare
photons with electrons in the forward shock; a subsequent interval
of slow decay would then be due to a progressive decline in this
suppression.  The range of acceptable models demonstrates that the
kinetic energy and circumburst density of GRB\,050904 are well above
the typical values found for low-redshift GRBs.

\end{abstract}

\keywords{gamma rays: bursts: individual: GRB050904 - cosmology: miscellaneous}

\section{INTRODUCTION}
\label{sec:intro}

One of the most exciting results from the first year of operations of
NASA's \swift\ satellite mission \citep{gcg+04} has been the discovery
and observation of GRB\,050904 \citep{cmc+06} at redshift $z=6.29$
\citep{kka+05}.  This burst was initially detected by the \swift\ BAT
at 01:51:44 UT on September 4, 2005, and was quickly followed up by
pointed observations with the \xray\ telescope (XRT) and UV/optical
telescope on \swift, and by numerous ground-based facilities.  Early
afterglow photometry provided the first indications for a very high
redshift for this event, $z>5.3$ \citep{gcn.3914}, prompting a global
observing campaign that culminated in the spectroscopic observations
by Subaru that provided the redshift \citep{kka+05} and enabled the
first investigation of the reionization epoch via GRB afterglow light
\citep{tkk+06}.

Within a day of the burst, the discovery of an associated prompt
$I\approx 14$\,mag optical/NIR flash was reported \citep{boer+05}.
The timing of this flare, which peaks at $t\approx 470$\,s after the
burst, is coincident with \xray\ (XRT) and $\gamma$-ray (BAT) flares
observed by \swift\ \citep{cmc+06b}.  This is the second such bright
optical flash observed so far, after GRB\,990123 \citep{abb+99}, and
may even exceed that event in optical luminosity \citep{kmk+06}.

Subsequent optical and NIR observations of the fading afterglow were
reported by \citet{hnr+06} and \citet{tac+05}, and provide evidence
for a slow decay phase during the first day after the burst, and a
jet break at $t\approx 3$\,days.  A campaign of radio observations
at the VLA yielded multiple detections of the afterglow at 8\,GHz,
which exhibited a slow evolution consistent with a high circumburst
density and extremely high kinetic energy \citep{fckn06}.  The
\xray\ lightcurve observed by \swift\ exhibits numerous interesting
features, including the early flare at $t\approx 470$\,s, vigorous
\xray\ flaring activity, and a possible jet break \citep{cmc+06b}.
Finally, observations with the \hstlong\ and the \spitzerlong\ by
\citet{bcc+06}\footnote{\citet{be+07} have re-observed the position
with \hstlong\ on July 22 UT, 2006 and then provide the upper limit
on the host galaxy. In the modeling, we have used the updated data
point.} have yielded late-time detections of the afterglow and host
galaxy.

Studies of GRBs at low-redshift, $1\siml z\siml 3$
\citep{pk+01,yhsf+03}, have established that the circumburst densities
for these events range between $0.06$ and $30\ {\rm cm^{-3}}$, with
their kinetic energies having a relatively narrow distribution with a
peak around $5\times 10^{50}\ {\rm ergs}$ \citep{pk+02,bkf03}. It is
therefore interesting to investigate whether the quantities for the
high-redshift GRBs follow these results or not.

Several questions seem particularly pertinent.  What density range is
found in the environments of GRBs at high redshift?  How do their
kinetic energies compare to those of low-redshift events?  Are other
properties of high-redshift GRBs -- their beaming angles and shock
microphysical parameters -- the same as for low-redshift GRBs?  The
answers to these questions can potentially cast light on outstanding
mysteries of the GRBs themselves, and reveal important aspects of the
early Universe.  These interesting questions provide the motivation
for our detailed investigation of the properties of GRB\,050904.

In this paper, we attempt a complete model of the full set of \xray,
optical/NIR, and radio afterglow observations of GRB\,050904.  Our
derivation of physical parameters from afterglow observations is
carried out in the context of the fireball model
(\citealt{mm+06}, and references therein).

We make explicit consideration of two scenarios for the origin of the
\xray\ and optical flares at $t\approx 470$\,s. Our first scenario (A)
attributes the flares to internal shocks or engine activity, and
excludes the flares from afterglow fits, with only later observations
considered.  This is consistent with the approach of \citet{wyf+05},
who argued on the basis of the fast decay of the flares that they
could not be due to reverse shock emission.  In our alternate scenario
(B), however, we attribute the flares to emission from the reverse
shock.  In order to accommodate their fast decay, we use a starting
time for the asymptotic Blandford-McKee solution which is near the
start of the flare, later than the burst trigger time, which serves to
flatten the post-flare decay. In this scenario, the optical flare
comes from synchrotron radiation in the reverse shock, and the \xray\
and $\gamma$-ray flares are produced by synchrotron self-Compton
scattering (SSC) of photons in the reverse shock region. A
cosmology where $H_0=71~{\rm km~s^{-1}~Mpc^{-1}},
\Omega_\Lambda=0.73,~ {\rm and}~\Omega_m=0.27$ is assumed in
calculating the luminosity distance $D_L$. For GRB\,050904 at
$z=6.29$, the luminosity distance is $D_L=1.93\times 10^{29}$\,cm.

Our model and its supporting analytical formulae are presented in
\S\ref{sec:obsandtheory}, with several derivations reserved for the
appendices.  Our numerical simulation procedure and the data set used
in our fits are described in \S\ref{sec:numsim}, while in
\S\ref{sec:analysisres} we analyze the results, including the $J$-band
light curve (\S\ref{sec:earlyag}), the radio light curve
(\S\ref{sec:radio_light}), and our ultimate energy and density
constraints (\S\ref{sec:energy_density}). A discussion of the results
and our conclusions are presented in \S\ref{sec:disc} and
\S\ref{sec:conclusions}, respectively.


\section{OBSERVATIONS AND THEORETICAL FRAMEWORK}
\label{sec:obsandtheory}


\subsection{Burst and Afterglow Observations}
The \swift\ BAT observations of the prompt emission give a
$\gamma$-ray duration of $T_{90}=225 \pm 10$ s, a spectrum with
power-law photon index $\Gamma=1.34 \pm 0.06$, and a fluence of $\sim
5.4 \times 10^{-6} \ {\rm erg \ cm^{-2}}$ \citep{cmc+06}. Given the
burst redshift of $z=6.29\pm 0.01$, the isotropic-equivalent gamma-ray
energy is $E_{\gamma,{\rm iso}} \sim 10^{54}\ {\rm ergs}$.  XRT
observations began 161\,s after the burst and continued for 10~days
after the burst trigger, overlapping with BAT observations for about
300\,s before the high energy emission faded below the BAT threshold
\citep{cmc+06}.

\begin{figure}[ht]
\centerline{\psfig{file=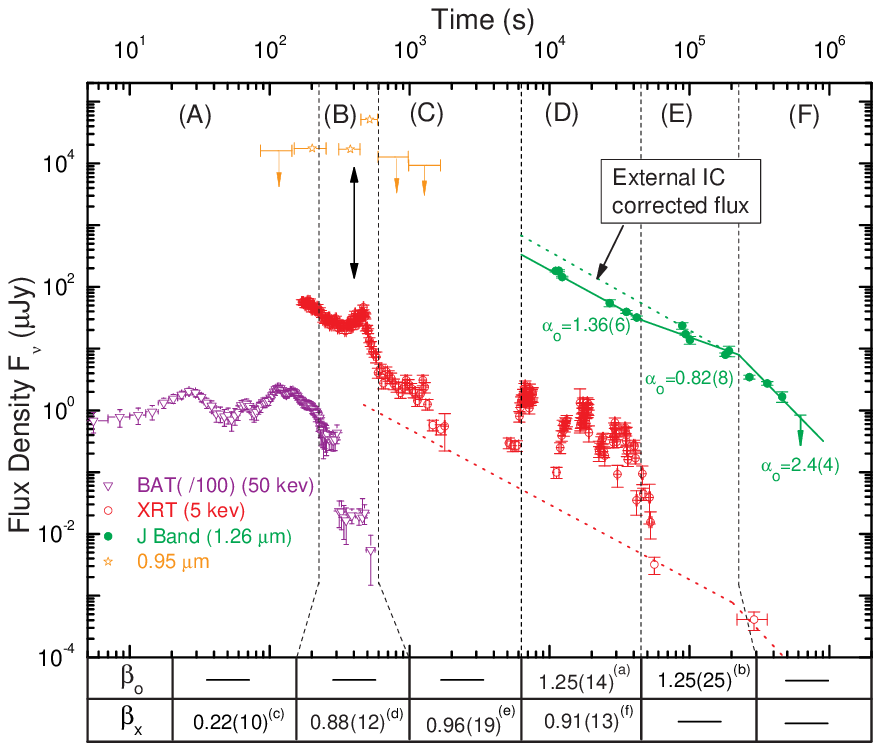,height=13cm}}
\caption{\small%
The combined light curves of GRB 050904 in the BAT, XRT, $J$ and $I$
bands, in the observer frame.  The BAT (violet, empty triangle) and
XRT (red, empty circle) data are taken from \citet{cmc+06}, the early
time $I$-band flare at 9500 $\rm \AA$ (orange, star) is from
\citet{boer+05}, and the $J$-band data (green, solid circles) are from
\citet{hnr+06} and \citet{tac+05}.  For ease of presentation, we show
the BAT flux density divided by a factor 100. The black arrows at
$t=398$ seconds after the burst indicate the adopted reference time
point, in our model (B), for the start of the afterglow evolution. The
solid line (green) shows the observed afterglow evolution in
$J$-band. The dashed lines show the theoretically-expected afterglow
evolution in the \xray\ and $J$-band, respectively, and the dash line
(green) is the light curve behavior in the J band without the external
IC process by the \xray\ photons from the flare.  The light curve is
divided into 6 sections labeled from A to F sequentially: (A) $t<
225$\,s: prompt emission; (B)$225 < t < 600$\,s: simultaneous flares
in the \xray\ and NIR bands; (C) $600 < t < 6220$\,s: power-law decay
in the \xray\ band; (D) $6220 < t < 4.3\times10^4$\,s: energetic
\xray\ flaring activity; (E) $0.5 {\rm days} < t < 1.6 {\rm days}$:
flattening of the light curve in the NIR band; (F) $t> 1.6$\,days: a
jet break is apparent.  Approximate temporal power-law indices
($F_\nu\sim t^{-\alpha}$) in the optical ($\alpha_o$) are noted.
Estimated spectral indices ($F_\nu\sim \nu^{-\beta}$) in the
optical/NIR (observed and uncorrected for extinction) and \xray\ are
also provided, where available, in the lower figure panels below the
figure. References for the spectral indices are (a) \citet{hnr+06};
(b) \citet{tac+05}; (c)-(f) \citet{cmc+06}.  A jet break time of
$t_{\rm jet} = 2.6\pm 1.0$\,days has been reported by \citet{tac+05};
however our fit of all available data suggests instead $t_{\rm jet} =
3.17 \pm 0.22$ days.}

\label{fig:all_the_data}
\end{figure}

Thanks to the \swift\ prompt alert, the early afterglow was also
observed promptly by the TAROT robotic telescope. The TAROT
observations started at 86\,s after trigger and lasted for more than
1500\,s; by making a spectrophotometric calibration of the field, Boer
et al. (2006) are able to present their data as flux densities at 9500
\AA, which we use and shall refer to as the TAROT $I$-band
observations.  Other larger ground-based telescopes started imaging
the field 3~hours later \citep{hnr+06,tac+05}.  We present our
compilation of the observational data in the \xray\ (XRT), hard \xray\
(BAT), and optical/NIR in Fig.~\ref{fig:all_the_data}. Here and
throughout the paper we convert \xray\ measurements to flux density
measurements at the frequencies of 5~keV (XRT) or 50~keV (BAT); these
energies correspond to observing frequencies of $1.2\times
10^{18}$\,Hz (XRT) and $1.2\times 10^{19}$\,Hz (BAT),
respectively. The conversion factors from photon counts per second
(cps) to flux are 3.31 $\mu{\rm Jy~cps^{-1}}$ for XRT PC data, 1.82
$\mu{\rm Jy~cps^{-1}}$ for XRT WT data, 154.6 $\mu{\rm Jy~cps^{-1}}$
and 86.2 $\mu{\rm Jy~cps^{-1}}$ for BAT masktag-lc data, respectively
(see \citeauthor{cmc+06}~\citeyear{cmc+06b} for the details of XRT PC,
WT and BAT data).

As seen in Fig.~\ref{fig:all_the_data}, the flux and spectral
evolution of the burst and afterglow divide the lightcurve into six
distinct segments, determined by inspection and motivated by the
physical model put forth herein. These segments are: (A) $t< 350$
seconds: In the \xray\ band, the flux decays as $F_{\nu} \propto
t^{-\alpha}$ with an index of $\alpha= 2.07 \pm 0.03$
\citep{cmc+06}. We follow the conventional definition for the flux
$F_{\nu} \propto t^{-\alpha} \nu^{-\beta}$. In the $I$-band
observation, there are two observational data points, the earlier of
which is only an upper limit, apparently indicating an increasing
tendency of the flux with time.  (B) $320 < t < 600$ seconds: Flares
are observed in both the $I$ and \xray\ bands, and also in the BAT
energy range.  They peak around 470 seconds after the burst trigger
time. The spectral index evolves from $0.50 \pm 0.07$ to $0.88 \pm
0.12$ over this time interval. (C) $600 < t < 6.3\times 10^3$ seconds:
A power-law decay is shown in the \xray\ band. During the same time
interval, there is no optical/NIR detection, except for two upper
limit flux values at 9500 ${\rm \AA}$. (D) $6.3\times 10^3 < t < 4.3
\times 10^4$ seconds: Many irregular fluctuations are observed in the
\xray\ band. The J band shows a decay which can be described with
$\alpha =1.36^{+0.07}_{-0.06}$ \citep{hnr+06}.  (E) 0.5 days $(\approx
4.3 \times 10^4\ {\rm seconds}) < t < 2.6 $ days: There is no
effective XRT observation within this period, and no further
fluctuations are detected. The flux in the J band is flattened a
little bit with a temporal index of $\alpha=0.82^{+0.21}_{-0.08}$
\citep{hnr+06} or $\alpha=0.72^{+0.15}_{-0.20}$ \citep{tac+05}. (F)
$t> 2.6$ days: the flux decay index in J band is around $\alpha=
2.4\pm 0.4$.  Only one data point in the \xray\ band is available. The
J-band data shows a sharp break, which is thought to be the jet break.


\subsection{Afterglow Modeling in the Swift Era}
\label{sec:swiftmodels}

Thanks to the rapid and precise alerts generated by \swift, and its
extensive multiwavelength follow-up campaigns, the afterglow data
collected during the \swift\ era has resulted in some necessary
modifications to the standard afterglow model.

Data from \swift\ have provided the greatest advance over earlier
datasets in the \xray\ band, as \swift\ responds to the burst orders
of magnitudes faster than previous generations of satellites and can
often track the \xray\ afterglow for up to 10 days. Many new features
of the \xray\ afterglow have thus emerged, leading to the
identification of a canonical X-ray afterglow behavior.  In addition
to the prompt emission phase, this involves at least five components
of the \xray\ afterglow (Nousek et al. 2006; Zhang et al. 2006a),
which are: (1) A steep decay phase, often interpreted as the tail of
the prompt emission, and thus, part of the GRB internal shock; (2) A
shallow decay phase of uncertain origin, with several theoretical
models proposed, including energy injection, jet inhomogeneities, or
varying shock microphysical parameters; (3) A normal decay phase,
familiar from pre-\swift\ observations; (4) A post jet break phase;
and (5) The \xray\ flares, which are superposed on the various
power-law segments of the afterglow's decay, and are fast-evolving in
the sense that the flare's rise and decay timescales $\delta t$ are
much smaller than the time since the burst $t$, that is, $\delta t/t
\ll 1$.  The current interpretation ascribes the \xray\ flares to the
same cause as the prompt gamma-ray emission -- energy dissipation in
internal shocks \citep{zfd+05}. Since the \xray\ flares are thus a
manifestation of central engine activity, it is necessary to exclude
them from afterglow model fits
\citep[e.g.,][]{fblc+06,cmrf+07,nggc+07}.

The accumulation of \swift\ afterglow observations has also raised
questions about the collimated or ``jet'' interpretation of afterglow
lightcurves.  Traditionally, jets have been invoked to explain a late
time ($t\simg 1.0$\,day), broadband and achromatic steepening in the
afterglow decay by an increment of +1 in the power-law index, from
$\alpha\approx 1.2$ to $\alpha\approx 2.2$.  Specifically, the
late-time decay index is predicted on robust grounds to be $\alpha=p$,
where $p$ is the power-law index of the energy distribution of the
synchrotron-emitting electrons that generate the afterglow light.

Current challenges to the jet break picture are two-fold.  First,
there are few jet breaks seen in the \xray\ light curves of afterglows
observed by \swift\ \citep{br+07}.  As of 7 October 2006, the XRT had
observed 145 long GRBs and 16 short GRBs, with the \xray\ afterglows
of almost all long bursts followed for days to weeks with \swift, in
campaigns that typically last until 10 days after the burst.
Unexpectedly, among these bursts, only 4 long bursts and 2 short
bursts show the jet break signature \citep[See Table 1,][]{br+07}.
Second, where \xray\ breaks of the appropriate steepness are seen,
they are often observed to be chromatic, exhibiting a different
evolution in UVOT (or ground-based optical) observations.  Thus the
jet break picture is being reevaluated, with new possibilities, such
as energy injection or the evolution of microphysical shock
parameters, under active consideration \citep{p05}.

Despite these results, we believe that the jet break picture remains
the better interpretation for most bursts, even in the \swift\ era.
First, the energetics of GRBs are extremely difficult to reproduce
from stellar-mass progenitors without beaming corrections of some
sort.  Second, the successes of the standard jet break picture in
modeling bursts from before \swift\ are too numerous and significant
to be discounted \citep[e.g.,][]{hbfs+99, pk+01b, yhsf+03}. Third,
several candidate jet breaks have been identified in the \swift\ era,
and when these are seen, they exhibit the expected properties
\citep[e.g.,][]{csff+07,dhma+07,br+07}.

Finally, several arguments support the presence of a jet break in the
evolution of the afterglow of interest to us here, GRB\,050904.  The
break that is observed in the NIR lightcurve of the afterglow at
$t=2.6\pm1.0$ day \citep{tac+05} is achromatic, being seen in multiple
NIR bands, and is followed by a steep post-break decay with power-law
index $\alpha=2.4\pm0.4$, consistent with the closure relation
expected from the standard jet model \citep{tac+05}.  Separately,
combining the Ghirlanda relation \citep{ggl+04}, and expression for
the jet break time \citep{sph+99}, a jet break is expected in the
range between 0.64 and 3.0 days.  The peak energy of GRB\,050904 is
$E^{src}_{peak}\ge 150(1+z)=1095$ keV and its isotropic-equivalent
gamma-ray energy lies between $6.6\times 10^{53}$ and $3.2 \times
10^{54}$ erg \citep{cmc+06}, so we take $E^{src}_{peak}\sim 1100$ keV
and $E_{\gamma,iso,52}\sim$ 100. In addition, the circumburst density
lies between $40< n < 1000\ {\rm cm^{-3}}$ \citep{kka+05}. Setting
$n\sim 100\ {\rm cm^{-3}}$ and the radiation efficiency in the prompt
emission phase $\eta_{\gamma}=0.2$, one can obtain the jet break
ranges of 0.64 and 3.0 days from Eqn. 3 of \citet{syis+07}, which is
consistent with the observed break time.


\subsection{Two Different Scenarios}

\subsubsection{(A) Forward Shock Fit Only}
It has been argued by \citet{wyf+05} that the flares at $t\approx 470$\,s are
due to internal shocks, rather than a reverse shock, based on an
apparent steep decay of the \xray\ light curve right after the peak,
with a temporal index of $\alpha_1\approx 8.8$ when referenced to the burst
trigger time (see \citealt{kz+06} for a discussion on the onset of GRB
afterglow). This is because the fastest decay must not be steeper than
that from the high latitude emission, whose temporal index is
$\alpha_2=\beta+2$, in this case, $\alpha\approx 3$.  This leads them to
favor internal shocks for both the optical and \xray\ flares, or a
combination of reverse shock for the optical and internal shocks for
the \xray. Consistent with this argument, we investigate a model of
the forward-shock emission only, that is, we consider only data points
in regions (C)-(F) of Fig.~1.

\subsubsection{(B) Forward Shock and Flares Fit}

Separately, we consider a scenario in which the flares peaking at
$t\approx 470$ seconds in the optical and \xray\ are due to the reverse
shock. This is motivated by the following argument. First, from the
observational point of view, the flares of GRB\,050904 have a great
similarity with the behavior of GRB\,990123, which is thought to be
due to the reverse shock (e.g., \citealt{sp99a}). Second, the
Blandford-McKee (BM) solution \citep{bm+76} describing the afterglow
is parameterized with a time origin $t_0$ given by the trigger time.
However, the BM solution is an asymptotic description of the afterglow
evolution, which is valid for times substantially longer than any of
the timescales associated with the prompt emission. Clearly, however,
there is a transition from the initial prompt phase, when the bulk
Lorentz factor is relatively constant, to the steady deceleration
phase when the asymptotic BM solution applies. Numerical simulations
have so far not provided specific guidance on the most appropriate
value of the reference time for BM (afterglow) evolution.

Naturally, however, the steepness of the light curve decay, as
parameterized by $F_\nu\sim (t-t_0)^{-\alpha}$, depends on the
reference time $t_0$ that is adopted. This has been discussed most
recently in the context of the \xray\ flares seen by \swift, which
exhibit steep decays and so are widely attributed to internal shocks
\citep{lzb+06}. Here we are dealing with a somewhat different
situation, in that the \xray\ light curve exhibits an initial flare
and a subsequent decay, which we propose to attribute to a reverse
shock. In the absence of guidance from numerical simulations, we adopt
a purely phenomenological approach, based on the constraint that,
whether for internal shocks or for a reverse external shock, the
temporal decay cannot be steeper than $(t-t_{0}^{'})^{-\alpha}$, where
$t_{0}^{'}$ is the reference time which fits the high latitude decay
$\alpha=2+\beta$, where $\beta$ is the spectral index. For late
central engine internal shocks, $t_{0}^{'}$ is found to be near the
onset of the last spike \citep{lzb+06}. Here, in the same spirit,
since the initial \xray\ flare is assumed to be due to the reverse
shock, after which the BM solution is asymptotically reached, the new
reference time $t_{0}^{'}$ is determined from the constraint that the
decay slope be equal to that expected from the high latitude
mechanism. Having set a new reference time $t_{0}^{'}$, all the
time-related quantities for the afterglow evolution, such as $\nu_m,
\nu_c$ and so on, are now referred to $(t-t_{0}^{'})$. We emphasize,
however, that although the reference time is shifted from the burst
trigger time to a new time point, the physical deceleration timescale
remains the usual one. For GRB\,050904, the best-fitting reference
time satisfying the high latitude condition is $t_{0}^{'}=(0.86\pm
0.01) t_{dec}$, where the (usual) deceleration timescale is 468\,s
(and the latter is measured relative to the burst trigger time).

The deceleration time is a critical point in the afterglow
evolution. After that, both the reverse shock and the forward shock
evolve into the asymptotic regime of the BM solution.  In model (B),
the peak time should represent the deceleration time of the shock. In
the discussion below all the formulae after the deceleration time are
taken relative to new reference time $t_{0}^{'}$ (we have inserted the
exact value of $t_{0}^{'}$ into the equations below). A difference
between models (A) and (B) lies in that $t_{0}^{'}=0$ for model (A)
and $t_{0}^{'}>0$ for model (B). The determination of $t_{0}^{'}$ for
model (B) will be described in \S\ref{sec:dataconst}.


\subsection{Synchrotron and Inverse Compton Afterglow Model}
\label{sec:ssc}

The flares in the optical \citep{boer+05} and \xray\ \citep{cmc+06}
are observed to peak simultaneously at $t\approx 470$\,seconds after the
burst. The \xray\ data has better time resolution, with an estimated
peak time of $t_{\rm peak}=468.0\pm 2.0$\,s.

We consider fits for two generic models, as described above.  In model
(A) the flares are ignored, as they may come from other regions
(possibly internal shocks) rather than the reverse shock region. In
model (B), however, we consider the flares to be due to the reverse
shock. In this case, since the peak of the flare is clearly separated
from the GRB itself, the reverse shock must be in the thin-shell
regime \citep{zkm+03}. Spectral analysis \citep{boer+05} shows that
the flares in the \xray\ and optical bands must be due to different
mechanisms. Here, we assume that the optical flare is due to
synchrotron radiation in reverse shock region, and that the \xray\
flare is produced by synchrotron-self-Compton (SSC)in the reverse
shock region (e.g., \citealt{wdl+01,kzmb+05}). Other observed \xray\
components at late times, except for the flares, are assumed to be
produced by synchrotron radiation from the forward shock.  We use the
following reference values for the main parameters involved: the
isotropic-equivalent kinetic energy $E_{52}=1$, the external circumstellar
density $n=1\ {\rm cm^{-3}}$, and the magnetic field ratio of the
reverse and forward shocks $R_B^{2} \equiv
\epsilon_{B,r}/\epsilon_{B,f}$. In addition, we have assumed equality
of the energy equipartition parameter for electrons in the forward and
reverse shock regions in model (B).

A transition from the temporal index $\alpha=0.82\pm0.08$ to
$\alpha=2.4\pm0.4$, interpreted as the jet break, is observed at
$t_{\rm jet}= 2.6 \pm 1.0$~days \citep{tac+05}, and the sharpness
of this break makes the burst not likely to have occurred in a stellar
wind-type density environment \citep{kp+00,gdhl+01,fckn06}. Therefore,
we will focus on afterglow evolution in a uniform density environment
throughout the paper.


\subsubsection{Forward Shock Synchrotron Formulae}
\label{sec:forsy}

Usually the synchrotron emission spectrum of the forward shock is
described by a broken power-law with three critical quantities
$\nu_m$, $\nu_c$, and $F_{\nu,{\rm max}}^{\rm syn}$ which are the
synchrotron frequencies for the electron energies at the injection,
cooling and peak flux emission, respectively \citep{spn+98}.  In
modeling the radio emission, it is necessary to also take into account
the self-absorption frequency, $\nu_{sa}$. We present the formulae
including self-absorption in Appendix \ref{self_ab}, because there are different
forms depending on various afterglow regimes, six in all, for the
spectra given by \citet{gs+02}. The main quantities of interest are:
\begin{eqnarray}
\nu_{m,f}&=&2.65 \times10^{23}\ {\rm Hz}\
\left(\frac{p-2}{p-1}\right) \epsilon_{B,f}^{1/2} \epsilon_e^{2}
\left(t-t_0^{'}\right)^{-3/2} \left(\frac{1+z}{7.29}\right)^{1/2}  \\
\nonumber \nu_{c,f}&=&1.58 \times10^{15}\ {\rm Hz}\ E_{52}^{-2/3}
\epsilon_{B,f}^{-3/2}
 n_{0}^{-1} \left(t-t_0^{'}\right)^{-1/2} \left(1+Y\right)^{-2}\left(\frac{1+z}{7.29}\right)^{-1/2}  \\
\nonumber F_{\nu,max,f}^{syn}&=& 7.63\times10^{-5}\ {\rm Jy}\
\epsilon_{B,f}^{1/2} E_{52} n_{0}^{1/2}
\left(\frac{D_{L}(z)}{1.9\times10^{29}}\right)^{-2}
\label{eqn:forward}
\end{eqnarray}
where the convention $Q=10^xQ_x$ is used, kinetic energy $E=10^{52}
E_{52}$~erg, and density $n=1\ n_{0}$~\percmcube.  $\epsilon_{B}$ is
the magnetic field equipartition parameter, and $\epsilon_e$ is the
electron equipartition parameter. The subscript ``$f$'' denotes the
forward shock, and ``$r$'' denotes the reverse shock. The parameter
$Y$ refers to the first-order inverse Compton effect and is defined as
$Y=(-1+\sqrt{1+\eta \epsilon_e/\epsilon_{B,f}})/2$ \citep{se+01}. Here
$\eta$, the fraction of the electron energy that is radiated away,
expresses the magnitude of the radiative correction: $\eta=1$ for fast
cooling and $\eta=(\gamma_c/\gamma_m)^{2-p}$ for slow cooling, where
$\gamma_c$ is the cooling Lorentz factor and $\gamma_m$ is the typical
Lorentz factor for the electrons. For $z=6.3$, the luminosity distance
in a concordance cosmology is $D_{L}\sim 1.9 \times
10^{29}$\,cm. After the deceleration time, relative to the new
reference time, the afterglow will evolve asymptotically as the BM
solution.  Thus, the evolution relation for each quantity is $\nu_c
\propto (t-t_{0}^{'})^{-1/2}, \nu_m \propto (t-t_{0}^{'})^{-3/2}$, and
$F_{\nu,m} \propto (t-t_{0,}^{'})^{0}$.

The quoted typical values for $\nu_m$, $\nu_c$ and $F_{\nu,{\rm
max},f}^{\rm syn}$ are consistent with the ones in \citet{gs+02}, but
the values for $\nu_m$ and $\nu_c$ are 2 times smaller, and the
typical value for $F_{\nu,{\rm max},f}^{\rm syn}$ is 4 times smaller
than that in \citet{spn+98}, which are fitted for the numerical
simulation results.

\subsubsection{Reverse Shock Synchrotron Formulae}
\label{sec:revsy}

In the reverse shock at the deceleration time, the typical quantities
are \citep{kzmb+05}:
\begin{equation}
\nu_{m,r}=\nu_{m,f} R_{B} R_{M}^{-2}, \ \nu_{c,r}= R_{B}^{-3} R_{X}^2 \nu_{c,f}, F_{\nu,max,r}=R_{B} R_{M} F_{\nu,max,f}
\end{equation}
where $R_{B} \equiv (\epsilon_{B,r}/\epsilon_{B,f})^{1/2}$,
$R_{X}\equiv (1+Y)/(1+X)$, $X=(-1+\sqrt{1+\eta
\epsilon_e/\epsilon_{B,r}})/2$ is the Compton parameter in the reverse
shock region, and the factor $R_{M} \equiv \Gamma_d^2/\Gamma_0$, where
$\Gamma_d$ is the Lorentz factor at deceleration time. For the
thin-shell case, $\Gamma_d \approx \Gamma_0$, so we have
$\Gamma_M\approx\Gamma_0$ \citep{zkm+03}.

After the deceleration time, since the shock is in the thin-shell
case, each quantity evolves as $\nu_c \propto
(t-t_{0}^{'})^{\alpha_c}$ where $\alpha_c={(15 g+24)/(14 g+7)}$,
$\nu_m \propto (t-t_{0}^{'})^{\alpha_m}$ where
$\alpha_m={-(15g+24)/(14g+7)}$, and $F_{\nu,m} \propto
(t-t_{0}^{'})^{\alpha_{f}}$ where $\alpha_{f}={-(11g+12)/[7(1+2g)]}$
\citep{zwd+05} where $g$ is the evolution index for $\Gamma \propto
R^{-g}$ where $\Gamma$ is the Lorentz factor of the afterglow, and
$R$ is the radius of the afterglow.

\subsubsection{Inverse Compton Effects, Jet Break and Non-relativistic Case}

Both models include inverse Compton effects. The formulae for the
normal case $\nu_a < \nu_m < \nu_c$ and $\nu_a < \nu_m < \nu_c$ in the
forward shock case are listed in \citet{se+01}. We have derived the
formulae for additional cases where the self-absorption frequency is
above the typical frequency $\nu_m$ or cooling frequency $\nu_c$ in
Appendix \ref{app:ic}. For inverse Compton effects in the reverse
shock, the self-absorption frequency has a similar form as that of the
forward shock, the only difference being that the
forward shock-related quantities are replaced by the corresponding
reverse-shock quantities.

Before the bulk Lorentz factor drops below the inverse of the opening
half-angle, $\theta^{-1}$, each of the typical quantities follows the
evolution given above (see sections \ref{sec:forsy} and
\ref{sec:revsy}). After $\Gamma < \theta^{-1}$, those typical
quantities follow a different evolution given by \citet{sph+99}.

We also consider the non-relativistic (NR) evolution of the afterglow
in its end-state. The time for the afterglow to enter into the NR
stage is calculated by the condition that the Lorentz factor of the
shell $\Gamma=2$. After the afterglow evolves into the NR stage, its
dynamics are described by the self-similar Sedov-von Neumann-Taylor
solution, for which \citet{fwk+00} have given a detailed description.

\subsubsection{Host Galaxy Extinction and Lyman-$\alpha$ Damping Absorption}

We have considered the host galaxy extinction in the optical band in
the rest frame of the host galaxy, using a Milky Way extinction curve.
We note that the particular type of extinction curve used is not expected
to make a difference in this case; \citet{kmk+06} have tested the
application of SMC, LMC, and Milky Way extinction curves to the
composite J-band data, and all three models suggest minimal
extinction. In our fitting, we treat the visual extinction $A_{V}$ in
the host galaxy as a free parameter. In addition, because the
extinction affects the spectral shape, we have considered the spectral
index correction due to the host galaxy extinction in our fitting.

Besides the normal extinction by the host galaxy, the emission close
to the wavelength of Lyman-$\alpha$ at $z=6.29$, including $z$ band
emission (effective wavelength $0.91\ \mu m$ and band width is $0.13\
\mu m$), has undergone neutral hydrogen absorption in the
IGM. \citet{tkk+06} fit the spectrum of GRB\,050904 and find its
best-fit column density to be $\log N_{HI} (\rm cm^{-2})=21.62$. To
find the expected Lyman-$\alpha$ absorption in the $z$ band, we have
convolved the Lyman-$\alpha$ absorption profile with the filter
transmission curve of the $z$-band, and obtain the absorption
coefficient $A=0.77$ meaning $33\%$ loss of $z$-band flux.

Also we notice that from Figure 6 in \citet{tkk+06}, the absorption
around wavelength $9500 {\rm \AA}$ is negligible, so we take the
absorption efficient $A=1$ for the 9500~\AA\ data.


\section{Fitting Data and Procedure}
\label{sec:numsim}


\subsection{Observational Data and Constraints}
\label{sec:dataconst}

Our model fits for the two different scenarios cover a range of bands
from the radio, through the IR/optical, to \xray\ and BAT
energies. The list of the observational data used in our global
fitting is in Table \ref{table:alpha-beta}.

\begin{table}[ht]
\caption{Reverse and Forward Shock Observation Data Points}
{\begin{tabular}{@{}lcccr@{.}l} \hline \hline \multicolumn{1}{c}{
No.} & \multicolumn{1}{c}{Obs. Time } & \multicolumn{1}{c}{Obs.
Freq.} &
\multicolumn{1}{c}{Band} &\multicolumn{2}{c}{Flux(Error)} \\
\multicolumn{1}{c}{} & \multicolumn{1}{c}{log10(t[sec.]) }  &
\multicolumn{1}{c}{log10($\nu$[Hz])} & \multicolumn{1}{c}{}
&\multicolumn{2}{c}{($\mu Jy$)} \\ \hline
1$\tablenotemark{\ddagger}$&2.66& 14.6 & $I$ & 5&1(10) $\times 10^4$ \\
2$\tablenotemark{\ddagger}$&2.66& 18.1 & 5 \kev\ & 70&4 (44) \\
3$\tablenotemark{\ddagger}$&2.66& 19.1 & 50 \kev\ & 4&45 (89) \\ \hline
4$\tablenotemark{\dagger}$ &2.86& 18.1 & 5 \kev\ & 0&65(26) \\
5$\tablenotemark{\dagger}$&3.25& 18.1 & -- & 0&32(7) \\
6$\tablenotemark{\dagger}$&4.75& 18.1 & -- & 0&32(10) $\times 10^{-2}$ \\
7$\tablenotemark{\dagger}$&5.46& 18.1 & -- & 0&51(17)$\times 10^{-3}$ \\
8&2.78&14.6 & $I$ & 12&8(64) $\times 10^3$ $\tablenotemark{\P}$ \\
9$\tablenotemark{\star}$ &4.98& 14.4 & $J$ & 19&1(73)\\
10& 5.27& 14.4 & -- & 9&2(17) \\
11& 5.56&14.4 & -- & 2&74(19) \\
12& 5.66&14.4 & -- & 1&67(32) \\
13& 5.79& 14.4 & -- & 0&42(21)$\tablenotemark{\P}$ \\
14 & 4.98 & 14.27 &  $H$ & 23&70 (21)\\
15 & 5.26 & 14.27 & --& 10&60 (10) \\
16 & 6.30 & 14.27 & -- & 8& 0(25) $\times 10^{-2}$ \\
17 &4.99& 14.14&  $K_s$ & 33&70(210) \\
18 & 5.27 & 14.14 & -- & 15&0(10) \\
19 & 4.94 & 14.51 & $z$ & 9&12(19) \\
20& 5.03 & 14.51 & -- & 6&92 (10) \\
21 & 4.97 & 14.46 & $Y$ & 14&0(30) \\
22 & 5.29 & 14.46 & -- & 8&4(22)\\
23$\tablenotemark{\clubsuit}$ &4.64& 9.9 & Radio &89&0(580) \\
24 & 5.08& 9.9 & -- & 41&0(250) \\
25 & 5.67&9.9 & -- & -3&0(250) \\
26 & 5.73& 9.9 & -- & 27&0(240) \\
27 & 6.24&9.9 & -- & 89&0(370) \\
28 & 6.40&9.9 & -- & 40&0(300) \\
29 & 6.46&9.9 & -- &-10&0(350) \\
30 & 6.47&9.9 & -- & 64&0(230) \\
31 & 6.48&9.9 & -- &116&0(180) \\
32 & 6.51&9.9 & -- & 67&0(170) \\
33 &6.58& 9.9 & -- &13&0(270) \\ \hline
\end{tabular}}
{\scriptsize
\tablenotetext{\ddagger}{The reverse shock emission. The NIR data are
from \citet{boer+05}, and the \xray\ data are from \citet{cmc+06}.}
\tablenotetext{\dagger}{The \xray\ data are from \citet{cmc+06}. For
the early-time \xray\ data, the contribution from the flares has been
subtracted.}  \tablenotetext{\star}{The $J$-band data are from
\citet{hnr+06} and \citet{tac+05}. Groups of adjacent data points have
been averaged.}  \tablenotetext{\clubsuit}{Radio data are from
\citet{fckn06}.}  \tablenotetext{\P}{Assuming both upper limits
as 2-$\sigma$ limits, we convert the upper limits into synthetic
measurements with error bars by taking the measurement to be half the
upper limit, and the error bar to be one-quarter the upper limit.}}
\label{table:alpha-beta}
\end{table}

In model (A), the new reference time is the trigger time, so $t_0=0$.
In model (B) with the flares included, we have to determine the new
reference time first. We will now describe our procedure for the early
time data used as input to the fits of model (B). Relative to the
trigger time, the temporal index of the fast decay is $\alpha\approx
8.8$. The fast decay is considered to be due to the high latitude
emission of the reverse shock. In order to obtain the new reference
time, we assume a simple power-law model for the high latitude flux
$f_{\nu} =F_{\nu}(\frac{t_{obs}-t_0}{t_{p}-t_{0}})^{-\alpha}$, where
$F_{\nu}$ denotes the peak flux at the peak time $t_{p}$, $t_{0}^{'}$
is the starting time at which the asymptotic Blandford \& McKee
solution applies, and $\alpha$ is the temporal index.  The high
latitude emission is assumed to decay with an index
$\alpha_{h}=2+\beta$. For the \xray\ observations right after the
peak, $t>470$ seconds after the burst, the observational spectral
index is around $\beta=0.88\pm 0.12$, so we take the temporal index in
the fitting as $2.88 \pm 0.12$, and selecting two observed data
points, we can find the new reference time to be $(0.86\pm 0.01)
t_{dec}$. In our fitting, we have set the new reference time at $0.86
t_{dec}$.

\begin{figure}[ht]
\plotone{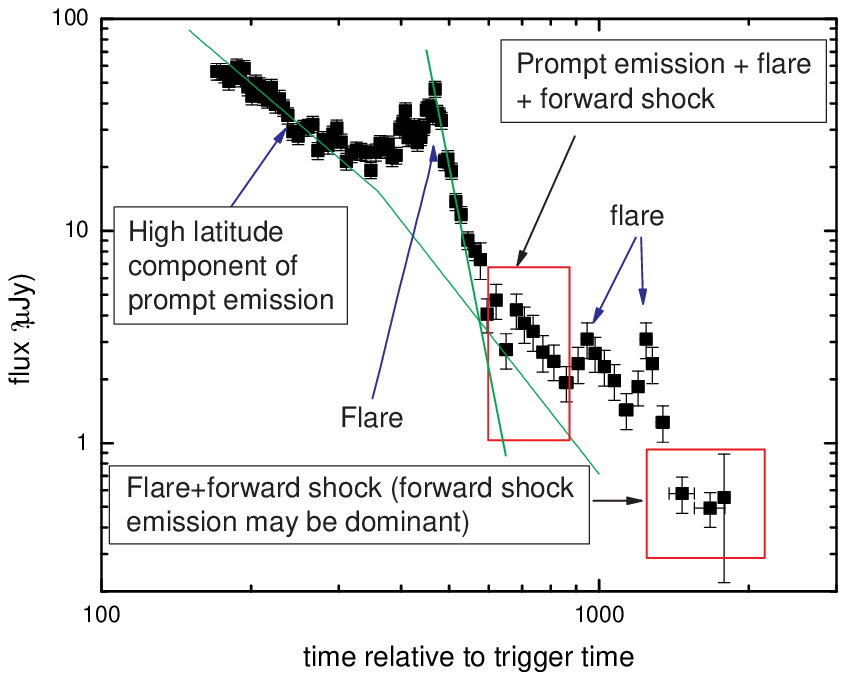}
\caption{\small Illustration of the multiple mechanisms contributing
to the observed flux at early times. The high-latitude component of
the prompt emission is described by a broken power-law with a break
time of $t\approx 350$\,s in the observer frame \citep{cmc+06}. Before
the break time, $\alpha\approx 2$ and afterwards, $\alpha\approx
3$. The main flare peaks at $t=468.0 \pm 2.0$\,s after the burst. The
black data points correspond to the observed flux. The green lines
indicate the theoretical extrapolation of the high-latitude emission
to late times, using the best-fit values of the burst and flare
emission without any subsequent fitting. The red-color boxes indicate
the data used for estimating the forward shock emission. }
\label{fig:flare_fitting}
\end{figure}

We notice that right after the fast decay starting 470~seconds after
the burst there is a plateau, extending from $t\approx 600$~seconds to
$t\approx 2000$~seconds (see Fig.~\ref{fig:flare_fitting}). A close
look at the data around 1000 seconds shows an obvious flare, so we
ignore the data after that. Thus, we only keep the \xray\ data between
470 seconds and 1000 seconds. For this part of the data, there are
three different contributions: (1) Forward shock; (2) High-latitude
emission of the flare; (3) High-latitude emission of the prompt
emission. It should be pointed that the high latitude emission of the
prompt emission must be described by a broken power-law due to the
spectral evolution of the burst itself \citep{cmc+06}; the break time
in the rest frame is at $t\approx 350$\,s in the observer
frame. Before the break time, the temporal index is $\alpha\approx 2$,
and afterwards, it is $\alpha \approx 3$.  We subtract the high
latitude emission from the observed flux to obtain the ``pure''
forward shock emission. To reduce the uncertainty of the data points,
we then find the mean flux by averaging adjacent data points in groups
of five, and use the averaged value for our fitting, finally providing
the fourth data point in Table~\ref{table:alpha-beta}.

We also apply a similar subtraction method to the data around $t\sim
2000$ seconds indicated in Fig.~\ref{fig:flare_fitting}. The averaged
data point is the 5th row in Table~\ref{table:alpha-beta}. One
difference between the subtraction for the fourth and fifth data
points is that we only consider the flare contribution for the fifth
data point (the prompt emission is considered negligible here), while
both the flare and the high latitude emission of the prompt emission
are considered for the fourth data point.

\begin{table}[ht]
\caption{Other observational constraints for the fitting}
{\begin{tabular}{clc}\hline \hline
 Constraint & Comments & Reference\\ \hline
$t_{\rm jet}=3.17\pm0.22^{\tablenotemark{1}}$ & Jet break time &1 \\
$t_{\rm dec}=468.0\pm2.0^{\tablenotemark{2}}$ & Deceleration time &2 \\
$\beta_{J}=1.25\pm0.25^{\tablenotemark{3}}$ &
      Spectral index in $J$ band at t=1.155 days &3\\
$\beta_{\rm X}=0.96\pm0.19^{\tablenotemark{4}}$ &
      Spectral index in \xray\ from t=680 and 1600 s. &4 \\
\hline
\end{tabular}}
\tablerefs{(1) The jet break is derived from fitting over all
available data in the \xray\ and NIR/optical. \citet{kmk+06} find
$t_{\rm jet}=2.63\pm 0.37$~days (1-$\sigma$), while \citet{tac+05}
found $t_{\rm jet}=2.6\pm1.0$~days. (2) See the text in
Sec.~\ref{sec:ssc}.  (3) \citet{tac+05}. (4) \citet{cmc+06}. }
\label{table:reverse-shock}
\end{table}

The common constraints considered in both model (A) and model (B) are
the following:
(1) The jet break time is fitted to be $3.17\pm 0.22$ days
(1-$\sigma$) by combining all available NIR and \xray\ data.
\citet{kmk+06} have extrapolated all the other NIR/optical data to the
J band and made a composite light curve in the J band, including the
HST data observed by \citet{bcc+06} $\sim 23$ days after the burst,
and obtain a jet break time $t_{\rm jet}=2.63\pm
0.37$~days. \citet{tac+05} give a value of $t_{\rm jet}=2.6 \pm
1.0$~days based on multiband fitting of a smaller data set.
(2) The spectral index in the $J$-band at 1.155 days is $1.25 \pm
0.25$ \citep{tac+05}.
(3) The average spectral index for the early time \xray\ afterglow
from $t=680$\,s to $t=1600$\,s is $\beta=0.96 \pm 0.19$
\citep{cmc+06}.

Besides the constraints above, we introduce another constraint for
model (B): (4) the deceleration time is $t_{\rm dec}=468.0 \pm 2.0$\,s
from our estimate of the peak of the \xray\ light curve.

It should be mentioned that in model (B) we have summed over both the
reverse and forward shock flux to fit the observed data ($I$, \xray\
and BAT bands) at the deceleration time.

Thus, considering all the available data listed in Table
\ref{table:alpha-beta}, as well as the observed spectral index and the
jet break time, we have 37 constraints for case (A) and 33 constraints
for case (B).

For the fitting and (simultaneous) evaluation of parameter
uncertainties, we perform a Markov Chain Monte Carlo analysis (MCMC;
see Sec.~\ref{sec:param} for a more detailed description).  In order
for the code to spend most of its time in the regions which have
physical meaning, we provide penalty conditions during the calculation
of the chi-square.  The penalty is levied by giving an extra boost to
the chi-square value when the penalty condition is violated (when the
condition is satisfied, this extra chi-square is zero). To ensure the
smoothness of the fit function at the penalty boundary, the penalty
function we choose (more or less arbitrarily) is $\Delta
\chi^{2}=[(x-x_{\rm lim})/(0.01*{\rm min(x, x_{\rm lim}}))]^4$, where
$x_{\rm lim}$ is the critical value for each parameter.

We have included some upper limits in our dataset by converting the
upper limits into synthetic measurements with error bars, as
follows. Assuming all the upper limits as 2-$\sigma$ limits (we notice
that the J band data labeled as No. 13 in Table \ref{table:alpha-beta}
is provided as a 3-$\sigma$ upper limit,
\citeauthor{tac+05}\,\citeyear{tac+05}; and the confidence level for
the I band data, No. 8, is not stated,
\citeauthor{boer+05}\,\citeyear{boer+05}), we take the measurement to
be half the upper limit, and the error bar to be one-quarter the upper
limit (half the synthetic measurement). For the radio data, which are
provided as measurements with error bars even when no detection is
realized, we adopt these measurements and error bars directly, while
plotting 2-$\sigma$ upper limits in our figures.

For both models (A) and (B), the parameter ranges are restricted as
follows: (1) $\epsilon_e \le 0.5$; (2) $\epsilon_{B,r} \le 0.5$; (3)
$\epsilon_{B,f} \le 0.5$; (4) Electron energy power-law index $p \ge
2.06$; (5) $\gamma_m>2.1$ at $t=10^7$ seconds after the burst.
Violations of the parameter ranges incur chi-square penalties as
discussed above, on a parameter by parameter basis.

Note that the critical electron energy index value (4) $p=2.06$ is
found by equating the minimum electron Lorentz factor for the $p>2$
and $p=2$ cases. For $p>2$, the typical Lorentz factor
$\gamma_m=\frac{(p-2)}{(p-1)} \frac{m_p}{m_e} \epsilon_e \Gamma$
\citep{spn+98}. For $p=2$, $\gamma_m \approx \frac{1}{\ln{(4.0\times
10^7)}}\frac{m_p}{m_e} \epsilon_e \Gamma$
\footnote[1]{The typical Lorentz factor for $p=2$ case can be obtained
in the similar way as the one for $p>2$}.

Also note that, because we have assumed that the critical radiation
mechanism is synchrotron radiation, the electrons in their shell frame
are relativistic, and correspondingly $\gamma_e >2$ or $\gamma_m>2$
(because at late time the afterglow are in the slow-cooling regime,
the Lorentz factor of the most electrons are concentrated at
$\gamma_m$). We have set the critical value to be (5) $\gamma_m=2.1$
rather than $\gamma_m=2$ to eliminate the unrealistic parameter set
corresponding to $\gamma_m \le 2$.

We have not constrained the extinction parameter $A_V$, since in
principle it can be any positive number; however, in order to avoid
considering negative values of the extinction, the code makes use only
of the absolute value of this quantity.


\subsection{Parameters and Methodology}
\label{sec:param}

For model (A), we have 7 free parameters, which vary freely, subject
only to the penalty conditions. The parameters are: the energy index
$p$, the isotropic-equivalent kinetic energy in units of $10^{52}$
ergs $E_{52}$, the energy fraction in electrons in the reverse shock
and forward shock regions $\epsilon_{e}$ (note that the electron
equipartition parameter in the forward shock is assumed to be the same
as the one in the reverse shock), the magnetic field equipartition
parameter in the forward shock $\epsilon_{B,f}$, the circumburst
density $n$, the jet opening half-angle $\theta$, and the extinction
parameter in the host galaxy $A_V$. In model (B), we introduce two
additional parameters: the magnetic field equipartition parameter in
the reverse shock $\epsilon_{B,r}$, and the initial Lorentz factor
$\Gamma_0$. We have assumed that the magnetic equipartition parameter
in the forward and reverse shocks can be different, which is motivated
by the results of \citet{fdhl+02,zkm+03,kp+03}.

In order to obtain best-fit parameters and explore the parameter space
of the fit function, we have tested both a grid search method and a
Markov chain Monte Carlo (MCMC) method.  However, we choose the MCMC
after some tests. The grid-based likelihood analysis calculates
chi-squared values at each grid point of the parameter space, and
determines the best fit parameters and confidence levels by finding
the minimum chi-square value point and range of values within a
certain ``height'' above that minimum. The benefit of this method is
primarily that it is straightforward. Once the parameter ranges and
the number of grid points are defined, the code is easily implemented.
However, the drawback is that it requires prohibitive amounts of time,
especially if there are many free parameters. For example, a coarse
grid with 7 points per dimension and with 8 parameters requires
$2.1\times 10^7$ evaluations, and at 0.2\,s per evaluation, the
calculation takes $\tau\approx 5$~days on a single processor
machine. Increasing the number of parameters, much less increasing the
number of grid points, quickly becomes infeasible. By contrast, the
MCMC method is very efficient, with execution time scaling linearly
with the number of parameters, which allows us to perform likelihood
analyses in a reasonable amount of time.

Briefly, the MCMC is a method to reproduce, directly, the posterior
distribution of the model parameters (for a detailed treatment in an
astronomical context, see \citealt{vps+03}). After a limited
``burn-in'' phase, it should generate a random draw from the posterior
distribution for most new function evaluations. From this sample, we
can then estimate all of the quantities of interest for the posterior
distributions (the mean, variance, and confidence levels). As
mentioned above, the MCMC method scales approximately linearly with
the number of parameters, allowing one to perform a likelihood
analysis in a reasonable amount of time for a large number of
parameters.  After an initial burn-in period, and assuming that
convergence of several chains can be established, all samples can be
thought of as coming from a stationary distribution. In other words,
the chain has no dependence on the starting location (although a good
choice of starting points and step size can accelerate the chain
convergence).

In implementing a MCMC approach, two key and interrelated questions
are: (1) At what point does the chain converge, that is, how fast does
the chain realize the target distribution?; and (2) Does the chain
provide good mixing, that is, has the chain covered all interesting
portions of the parameter space?  \citet{gr+92} suggest a method to
test the convergence and mixing and introduce for this purpose a
parameter labeled $\hat{R}$ (see also \citealt{vps+03}). The
convergence can be monitored by calculating $\hat{R}$ for all the
parameters in two or more chains, and running the simulations until
all $\hat{R}$ values are less than 1.2. More conservatively, we may
choose to run until all $\hat{R} <1.1$; this is the criterion we adopt
as our test of convergence.

As we mentioned above, the MCMC model efficiently explores the
parameter space, which guarantees that the global minimum will be
approached in the long run. By contrast, in a grid search method,
one has to provide the parameter range beforehand, and one is never
sure that the minimum chi-square found is the global minimum instead
of a local minimum. This can lead to very different results.

Before running the MCMC, it is necessary to initialize the starting
point and assign step sizes for each parameter. For the starting
point, we run a test chain first, then choose the best parameter set
(evaluated by a minimum chi-square) as the starting point for a
formal run. Initially, we set the step size for each parameter to be
the 1-$\sigma$ range from this same initial run; however, several
experiments convinced us that a half-sigma step size provided better
convergence speed. Our MCMC code was implemented in a Matlab
environment on a single processor machine, and then transplanted to
the High Performance Computing (HPC) Linux cluster at Pennsylvania
State University. We made use of 4 processors, each running one
chain, and with each chain set to run for 2 million steps at a time.
After the chain calculations are completed, we merge them and test
for convergence. If the chains have not converged over the final
1~million steps of the 2~million-step chains, then we take the final
parameter set as the starting point for another run, execute
2~million additional steps, and test for convergence again. The
result for our final model, presented here, provided convergence
after the first run in both cases (Models A and B), with
$\hat{R}<1.1$ for all parameters after 2~million steps.


\subsection{Numerical Results}
\label{sec:numres}

For the results presented here, we ran four chains for $2\times
10^{6}$ steps for each of the models (A) and (B).  Convergence was
tested and parameters quantified using the final one million steps
only.  We found that both chains had already converged after the first
run.  For model (A) the $\hat{R}$ values are 1.008, 1.068, 1.045,
1.022, 1.076, 1.070, 1.004 for the parameters $p$, $\epsilon_e$
$\epsilon_{B,f}$ $n$, $\theta$, $E_{52}$, $A_V$, respectively. For
model (B), the $\hat{R}$ values are 1.002, 1.004, 1.002, 1.004, 1.005,
1.005, 1.000, 1.004, 1.000 for the parameters $p$, $\epsilon_e$
$\epsilon_{B,f}$ $n$, $\Gamma_0$, $\theta$, $\epsilon_{B,r}$,
$E_{52}$, $A_V$, respectively.

\begin{figure}[ht]
\centerline{\psfig{file=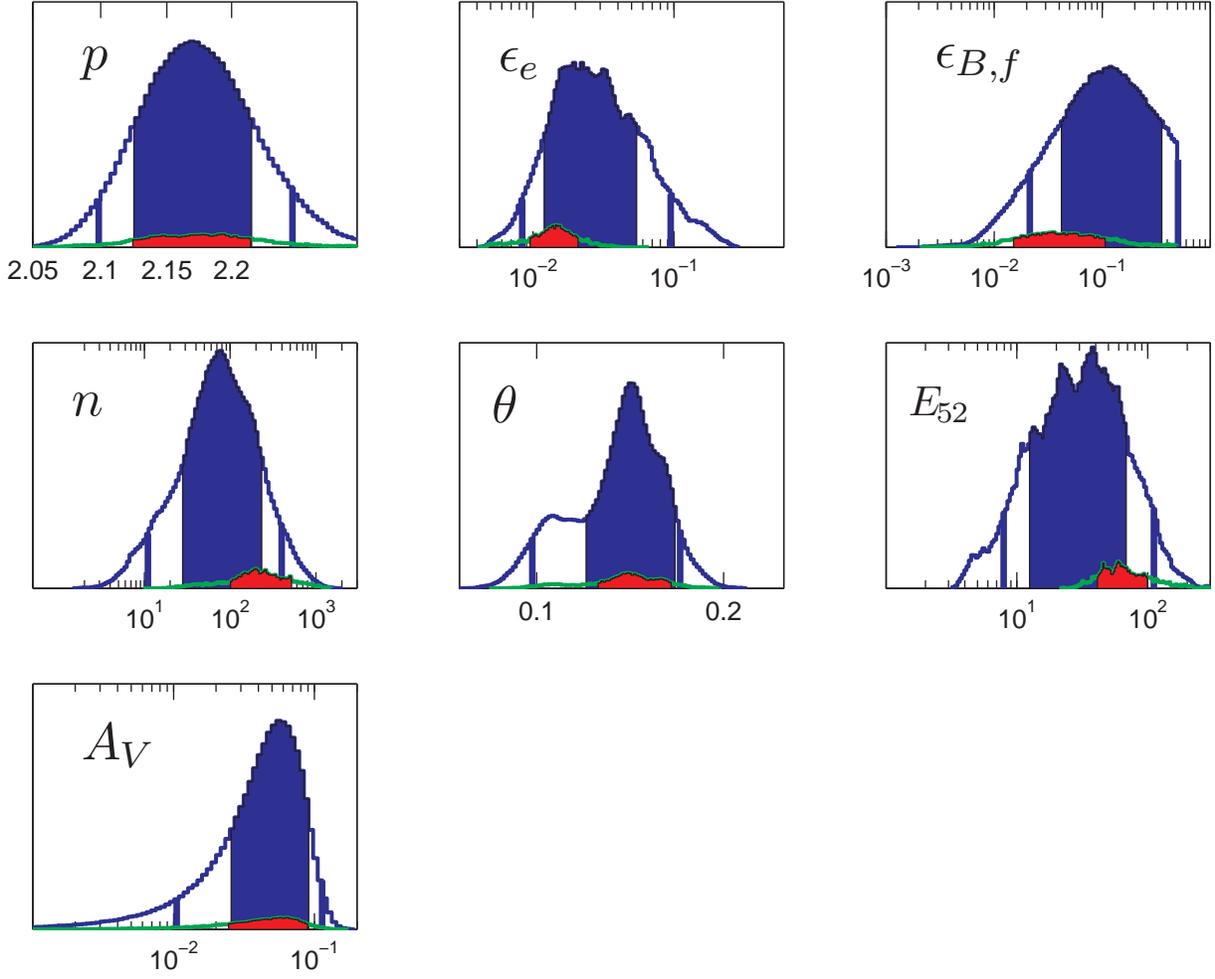,height=13cm}}
\caption{\small%
Posterior distribution of all parameters for model (A), which excludes
data related to the flaring activity at $t\approx 470$\,s: $p$, the
power-law index of the electron energy distribution; $\epsilon_e$, the
electron equipartition parameter; $\epsilon_{B,f}$, the magnetic field
equipartition parameter in the forward shock; $n$, the circumstellar
density in units of $\rm cm^{-3}$; $\theta$, the opening half-angle before
jet break; $E_{52}$, the isotropic-equivalent kinetic energy in units of
$10^{52}$ ergs; and $A_{V}$, the dust extinction of the host
galaxy. In each plot, the shaded blue region delimits the 1-$\sigma$
(68.2\%) confidence range, and the vertical lines indicate the
90\%-confidence range.  The green color indicates the posterior
distribution of the parameters for models having $\gamma_m<2.1$ at
$t=10^7$\,s, with the red lines indicating 1-$\sigma$ confidence
ranges, and the height of the distribution magnified by a factor of
five for clarity; these model realizations correspond to roughly 1.5\%
of the total.}
\label{fig:for_dis}
\end{figure}

The posterior distributions for the parameters for model (A) are
displayed in Fig.~\ref{fig:for_dis}. The shaded blue regions delimit
the 1-$\sigma$ range (68.2\%-confidence), and the region included
within vertical blue lines corresponds to the 90\%-confidence
interval. We choose the ranges of minimum width for these confidence
intervals.  Separately, we indicate the posterior distribution of model
parameters for models with $\gamma_m<2.1$ (at $t=10^7$\,s) in green,
with a red color for its $1\sigma$ region. The number of model
realizations that were constrained in this sense is roughly $1.5\%$ of
the total trials. The reduced chi-square for model (A) reaches its
minimum value $36.2/26 \approx 1.39$, and the best fit parameters are
$p=2.152$, $\epsilon_e=0.031$, $\epsilon_{B,f}=0.198$,
$n=84.4$~\percmcube, $\theta=0.128$, $E_{52}=22.4$, and $A_V=0.034$\,mag.

\begin{figure}[ht]
\centerline{\psfig{file=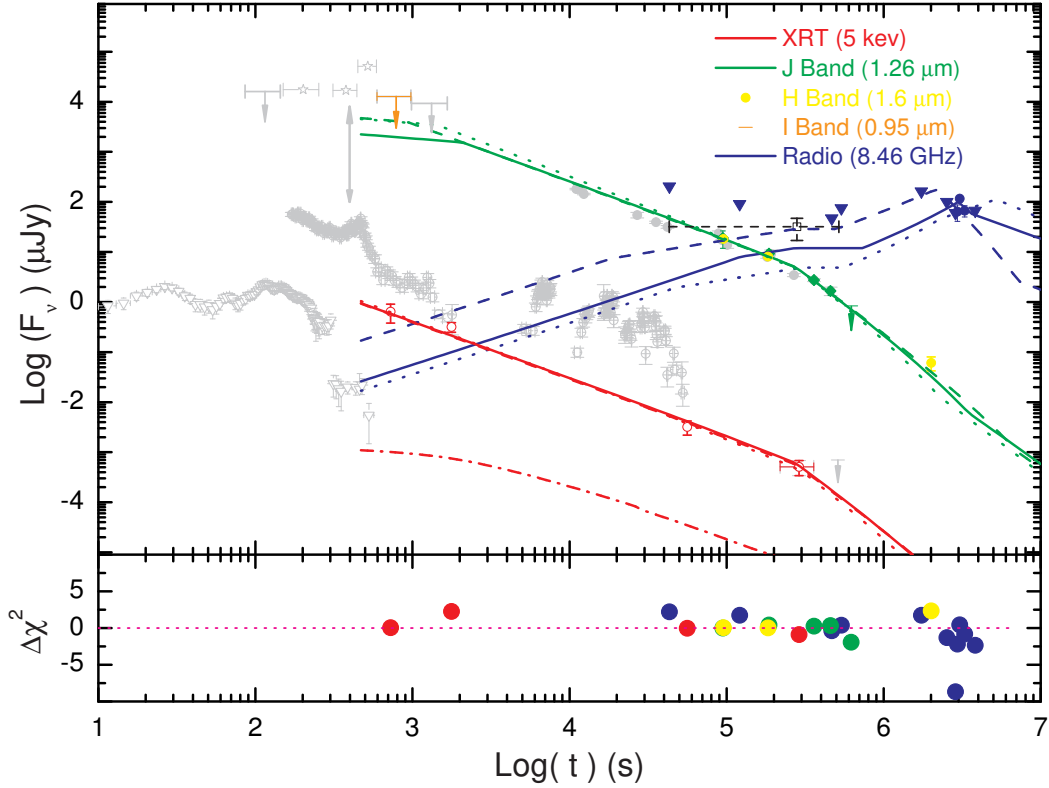,height=13cm}}
\caption{\small%
{\bf Top}: Theoretical light curves (solid lines) corresponding to the
best-fit parameters for model (A), which excludes data related to the
flaring activity at $t\approx 470$\,s.  Best-fit model parameters are
$p=2.152, \epsilon_e=0.031, \epsilon_{B,f}=0.198, n=84.4\,{\rm
cm^{-3}}, \theta=0.128, E_{52}=22.4, A_V=0.0343$ mag.  We have
shown all the available data on the plot.  If the data point is used
in the fitting, it is plotted with a bright color, otherwise it is
plotted in grey.  To illustrate the effects of different densities for
model (A), we show additional light curves. The dashed lines
correspond to $n=10^3 \ {\rm cm^{-3}}$, with marginalized best-fit
parameters of $p=2.25,
\epsilon_e=0.019,\epsilon_{B,f}=0.043,\theta=0.182, E_{52}=44,
A_V=0.0426$ mag, and with reduced chi-squared value $46.1/26 =1.77$.
The dotted lines correspond to $n=10\ {\rm cm^{-3}}$, with
marginalized best-fit parameters of $p=2.19$, $\epsilon_e=0.014$,
$\epsilon_{B,f}=0.078$, $\theta=0.098$, $E_{52}=50.5$,
$A_V=0.064$\,mag, and reduced chi-squared value $42.1/26 =1.62$.  All
light curves show the total of synchrotron and inverse Compton
emission; optical and near-infrared data have been converted to
$J$-band flux densities for clarity in plotting.  For the best-fit
model only, we plot the contribution to the \xray\ flux from inverse
Compton emission separately, as the red dash-dotted line.
{\bf Bottom:} The chi-squared contribution from each data
point. Positive values indicate that the best-fit model underestimates
the flux, and negative values indicate that it overestimates the
flux.}
\label{fig:light_curve_density}
\end{figure}

The light curve for the best fitting parameters is shown in
Fig.~\ref{fig:light_curve_density} (indicated with solid line). We
have shown the observational data as the background in the grey color,
and indicated the data actually used for fitting in other bright
colors (for detailed description on which color stands for which data,
refer to the caption of the figure). We have plotted all of the data
used, except for $Y$-band, $z$-band, and $K_s$-band data points, which
may overlap with $J$-band data points if these non-J-band data are
converted to $J$-band. For clarity of presentation, we convert
non-$J$-band NIR data to $J$-band on the basis of the spectral index
at that point, but still label it with its original band name and plot
in a different color. At the bottom of the light curve figure, we
show, as ``residuals,'' the chi-square contribution of each data
point. We notice that the radio data provide a large contribution to
the total chi-square, because of unexpected variations from
observation to observation which are difficult to reproduce.  In
particular, the model fails to remain within the several upper limits
from radio observations (note that eight out of 11 radio observations
are upper limits, and three are flux measurements).

Fig.~\ref{fig:light_curve_density} also demonstrates the effects of
the density on the afterglow evolution (shown in dashed and dotted
lines, respectively). We notice that the \xray\ and optical/NIR can be
fitted well even for density values varying by 2 orders of magnitude;
the only significant effects are seen in the radio light curve. The
radio lightcurve for a density $n=10$~\percmcube\ medium peaks around
$t\sim 2.0\times 10^6$\,s at a flux of $F_\nu\approx 200$\,\uJy, that
for a density $n=84.4$~\percmcube medium peaks around $t \sim 3.0 \times
10^{6}$\,s at $F_\nu\approx 100$\,\uJy, and that for a density
$n=10^3$~\percmcube\ medium comes even later, at $t\sim 5.0 \times
10^6$\,s and $F_\nu\approx 120$\,\uJy. At this point, the afterglow is
in the regime with $\nu_m < \nu_a < \nu_c$, and the peak flux is at
the self-absorption frequency.  Because the self-absorption frequency
is a function of density, the larger the density, the larger the
self-absorption frequency. This predicts an earlier peak for a lower
density. With more observational data points, or even upper limits
after $t \sim 3\times 10^6$\,s, we could put stronger constraints on
the circumstellar density.

\begin{figure}[ht]
\centerline{\psfig{file=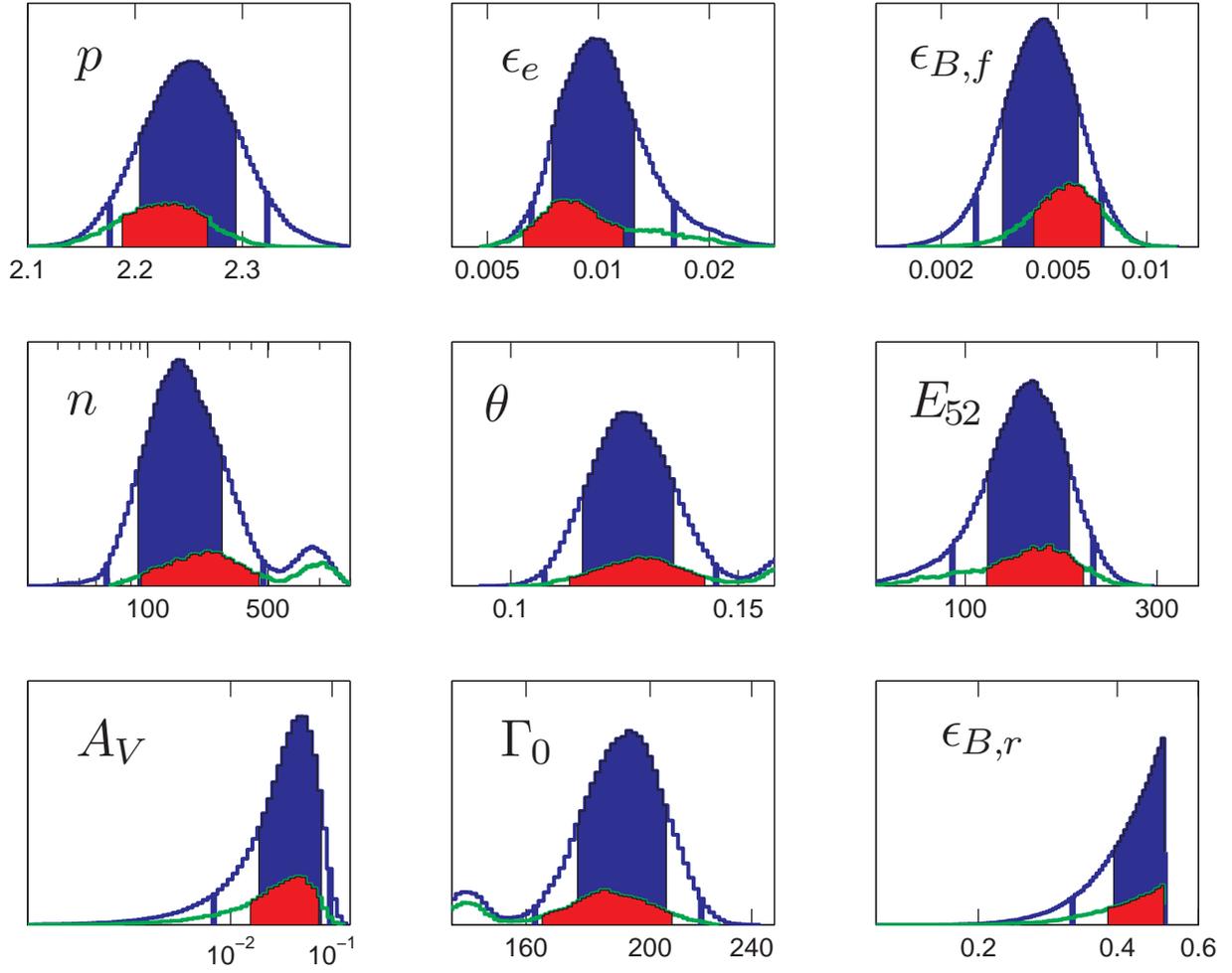,height=13cm}}
\caption{\small%
Posterior distribution of all parameters in model (B), in which the
flares at $t\approx 470$\,s are considered to be emission from the
reverse shock regions: $p$, the power-law index of the electron energy
distribution; $\epsilon_e$, the electron equipartition parameter;
$\epsilon_{B,f}$, the magnetic field equipartition parameter in the
forward shock; $n$, the circumstellar density in units of $\rm
cm^{-3}$; $\theta$, the opening half-angle before jet break; $E_{52}$,
the isotropic-equivalent kinetic energy in units of $10^{52}$ ergs;
$A_{V}$, the dust extinction of the host galaxy; $\Gamma_0$, the
initial Lorentz factor of the outflow; and $\epsilon_{B,r}$, the
magnetic field equipartition parameter in the reverse shock. The
shaded blue region delimits the 1-$\sigma$ range ($68.2\%$), and the
region included within the vertical blue lines corresponds to
$90\%$-confidence level. The green color indicates the region for
$\gamma_m<2.1$ (the red color corresponds to 1-$\sigma$ confidence
range), which is around $4.7\%$ of the total trials; ; its height has
been magnified by a factor of five for clarity.}
\label{fig:for_rev_dis}
\end{figure}

The posterior distributions for the parameters for model (B) are shown
in Fig. \ref{fig:for_rev_dis}. As for model (A), most of the
distributions are satisfyingly Gaussian in shape. One exception is the
magnetic field equipartition parameter $\epsilon_{B,r}$, seen to peak
at $\epsilon_{B,r}=1/2$, the upper bound for $\epsilon_{B,r}$ in our
model.  Also, we note that the posterior distributions for the density
$n$, the initial Lorentz factor $\Gamma_0$, and the opening half-angle
$\theta$, have irregular tails. These irregular tail regions
correspond to a distinct (local) chi-square minimum.

\begin{figure}[ht]
\centerline{\psfig{file=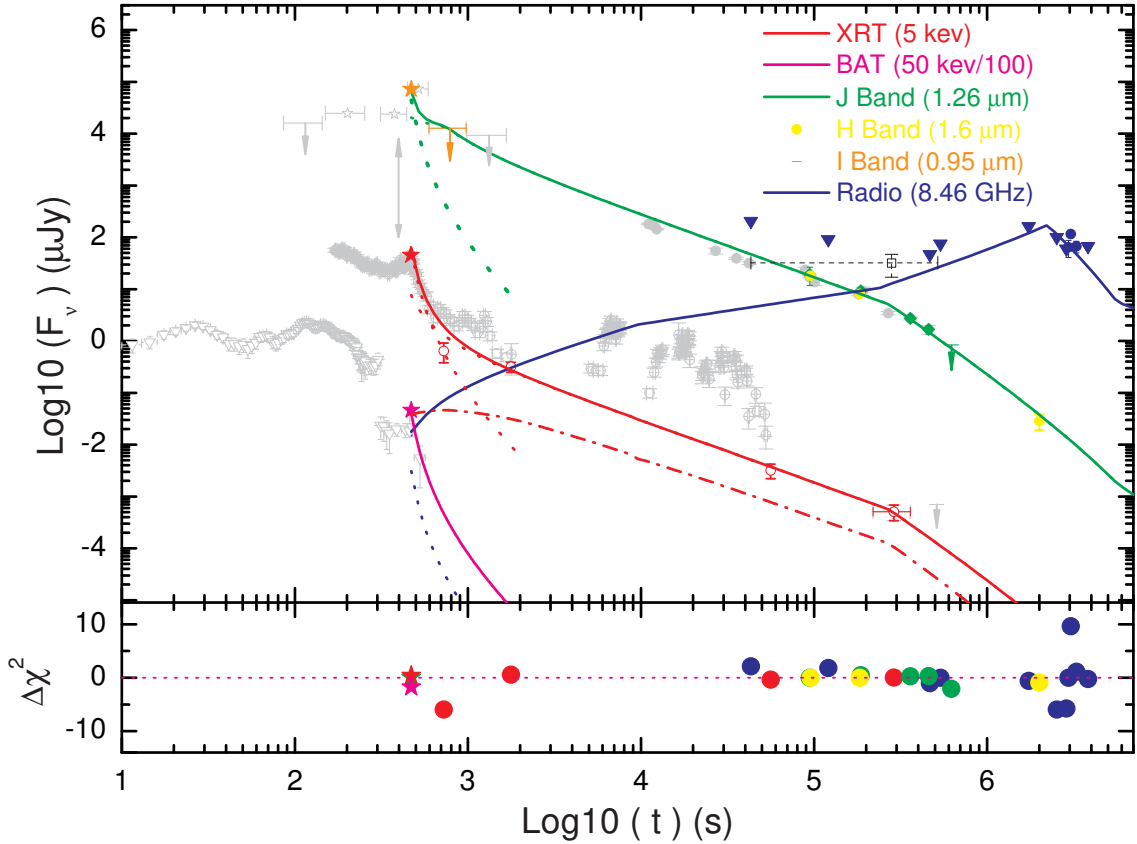,height=13cm}}
\caption{\small
{\bf Top:} Theoretical light curves (solid lines) corresponding to the
best-fit parameters in model (B), in which the flares at $t\approx
470$\,s are considered to be emission from the reverse shock regions.
Best-fit model parameters are: $p=2.243$, $\epsilon_e=0.0084$,
$\epsilon_{B,f}=5.7\times 10^{-3}$, $n=212.4\,{\rm cm^{-3}}$,
$\theta=0.126$, $E_{52}=147$, $A_V=3.18 \times 10^{-2}$\,mag,
$\Gamma_0=183.6$, and $\epsilon_{B,r}=0.50$. The dotted lines indicate
the separate flux contributions from the reverse and forward shocks
(reverse shock emission is distinguished by its fast decay at early
times).  The solid lines indicate the total model flux, with the red
dash-dotted line showing the contribution to the \xray\ flux from
inverse Compton emission, which is relatively unimportant compared to
the synchrotron component.  All optical and near-infrared data have
been converted to $J$-band flux densities for clarity in plotting;
underlying model calculations use the various observed frequencies
directly.  Radio, $I$-band, $J$-band, $H$-band, \xray, and BAT data
are plotted in blue, orange, green, yellow, red, and magenta,
respectively, as indicated in the plot legend.
{\bf Bottom:} The contribution to the total (best-fit) model chi-squared
for each data point.  Positive values indicate that the model
underestimates the flux, and negative values indicate that it
overestimates the flux.}
\label{fig:light_curve_flare}
\end{figure}

In the region where the reduced chi-square for model (B) reaches its
minimum value $53.0/28 \approx 1.9$, the best fit parameters are
$p=2.243$, $\epsilon_e=0.0084$, $\epsilon_{B,f}=5.7 \times 10^{-3}$,
$n=212.4$~\percmcube, $\theta=0.126$, $E_{52}=146$, $A_V=0.032$\,mag,
$\Gamma_0=183.6$, and $\epsilon_{B,r}=0.50$.  The light curve for
these parameters is given in Fig. \ref{fig:light_curve_flare}.

\begin{figure}[ht]
\centerline{\psfig{file=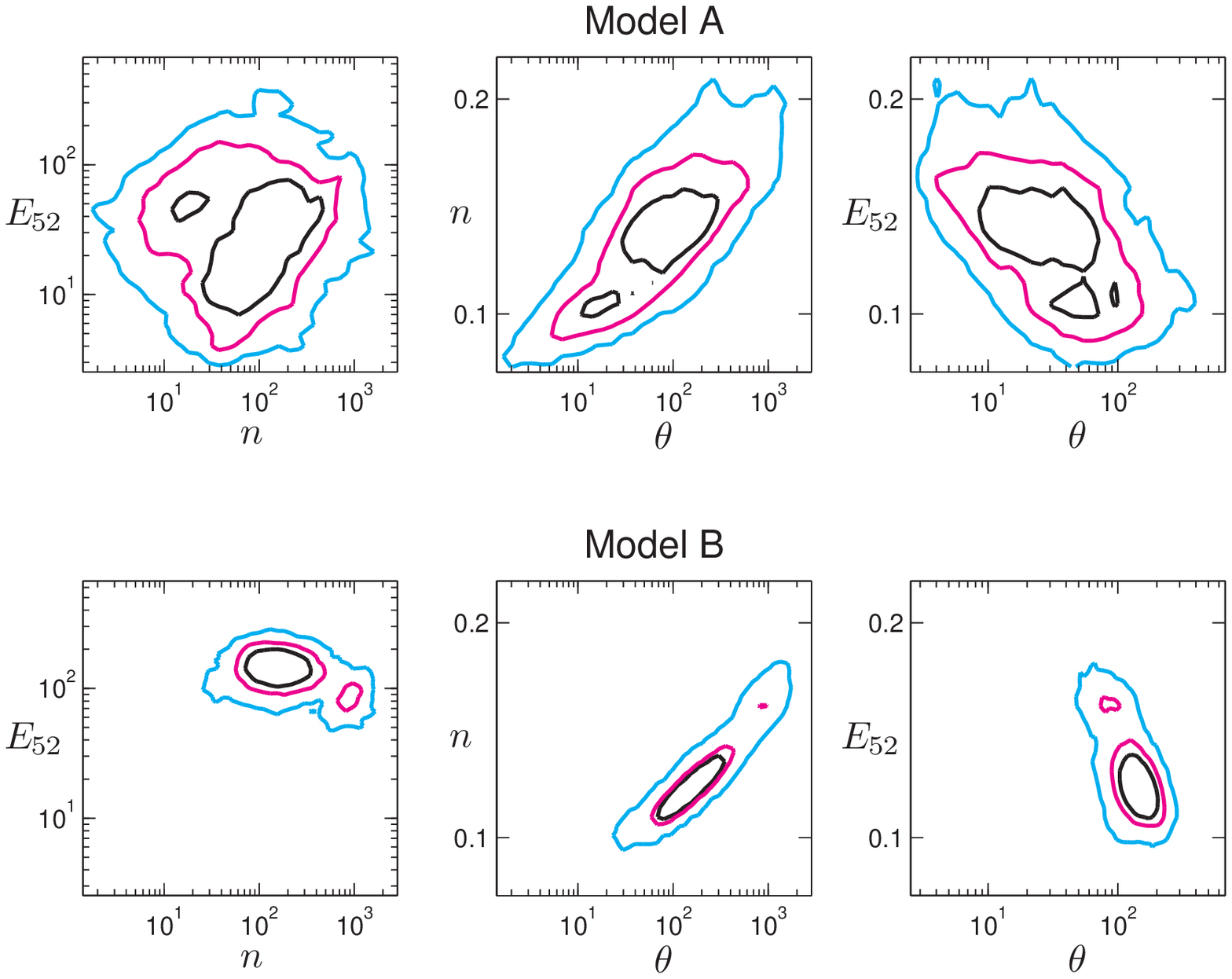,height=13cm}}
\caption{\small
Joint confidence regions ($68\%, 90\%, {\rm and}~99\%$, respectively) for
three chosen parameters from our model fits: the jet opening
half-angle $\theta$, the isotropic-equivalent kinetic energy
$E_{52}$, and the circumburst density $n$.
{\bf Top:} Joint confidence regions for model (A), which excludes data
related to the flaring activity at $t\approx 470$\,s.
{\bf Bottom:} Joint confidence regions for model (B), in which the
flares at $t\approx 470$\,s are considered to be emission from the
reverse shock regions.  With the additional constraints available in
model (B), the separate effects of blastwave kinetic energy and
circumburst density can be distinguished, and the final constraints on
the beaming of the burst and its total kinetic energy are
significantly better-defined.}
\label{fig:contour_plot}
\end{figure}

In Figure \ref{fig:contour_plot}, we show contour plots for the
joint confidence regions of three important physical parameters: the
jet opening half-angle $\theta$, the isotropic-equivalent kinetic
energy $E_{52}$, and the circumburst density $n$. This illustrates
the degree of covariance between these quantities, in a quantitative
manner.

In Table \ref{table:parameter_value} we list the best-fitting
parameters, and the parameter ranges for 1-$\sigma$ and $90\%$
confidence level for both models (A) and (B).

\begin{table}[ht]
\caption{The best fit values and parameter ranges for models (A) and
(B).} {\begin{tabular}{@{}ccccccc}\hline\hline &
\multicolumn{3}{c}{(A) Forward shock only} & \multicolumn{3}{c}{(B) Reverse shock flare} \\
\cline{2-7} \raisebox{1.5ex}[0pt]{Parameters} & Best Fit & 1 $\sigma$
Range & $90 \%$ Range & Best Fit & 1 $\sigma$ Range & $90 \%$ Range\\
\hline $p$& 2.15 & 2.11--2.19 & 2.09--2.22 & 2.24 &
2.20--2.29 & 2.18--2.32 \\ $\epsilon_e(/10^{-2})$ & 3.09 &
4.3--14.6 & 2.8--26.3 & 0.84 & 0.75--1.3 & 0.66--1.6\\
$\epsilon_{B,f}(/10^{-2}$) & 19.8 & 4.5--38.9 & 2.0--50.5 & 0.57 &
0.32--0.58 & 0.26--0.70 \\ $n$ & 84.4 & 26--273 & 9 -- 580
& 212.4 & 88--271 & 58--470 \\ $\theta$ & 0.128 &
0.12--0.18 & 0.10--0.19 & 0.126 & 0.11--0.13 & 0.11--0.14 \\
$E_{52}$ & 22.4 & 13--53 & 7--102 & 146.6 &
114--182 & 93--208  \\ $A_V (/10^{-2})$& 3.43 & 1.8--8.0
&0.7--10.6 & 3.18 & 1.9--7.8 & 0.7--9.6 \\ $\Gamma_0$&\ldots
&\ldots&\ldots& 183.6 & 176--206 & 163--219\\
$\epsilon_{B,r}$&\ldots &\ldots&\ldots& 0.50 & 0.4--0.5 &
0.3--0.5 \\  $\chi^2/{\rm dof}$& 36.2/26
&\ldots&\ldots&53.0/28&\ldots&\ldots \\ \hline
\end{tabular}}
\label{table:parameter_value}
\end{table}

\section{ANALYSIS OF THE RESULTS}
\label{sec:analysisres}

\subsection{$J$-Band Light-Curve and IC Suppression}
\label{sec:earlyag}

The shift of the afterglow starting time to the epoch $t_{0}^{'} \sim
402$ seconds modifies the analysis and the interpretation of the early
afterglow in model (B), but this should not affect the late-time
afterglow light curve evolution, which should be similar to the one
for model (A). Thus the discussion below will be focused on the light
curve of model (A), and where there are differences, these are pointed
out.

\citet{hnr+06} have fitted their collection of NIR data, and find that
between 3~hours and 0.5~days after the burst, the fading of the
afterglow can be fitted by a power law of index
$\alpha=1.36^{+0.07}_{-0.06}$, while after 0.5 days the fading appears
to slow down to a temporal index of $0.82^{+0.21}_{-0.08}$. At
$t=10.6$~hours (0.44 days), the spectral index is $\beta_o =
1.25^{+0.15}_{-0.14}$. A single power-law decay is ruled out at
3.7$\sigma$ confidence.  \citet{tac+05} have extended datasets whose
observation time reaches up to 7~days after the burst, and fit the
lightcurve with a smoothly-broken power law. The fit gives $\alpha_1
=0.72 ^{+0.15}_{-0.20}$, $\alpha_2=2.4\pm0.4$, and $t_b =2.6 \pm
1.0$~days.  In addition, the spectral index at $t=1.155$~days is
calculated to be $\beta=1.25\pm 0.25$ or $\beta=1.2\pm 0.3$ by two
slightly different fitting codes. Thus, based only on the observations
in the optical/NIR bands, we can divide the light curve into 3
segments: (D) $t<0.5\ {\rm days}$: the afterglow decays as a power-law
with index of $1.36^{+0.07}_{-0.06}$; (E) $ 0.5\ {\rm days} < t < 1.6\
{\rm days}$: the light curve is relatively flat, decaying as a
power-law with index $\alpha\approx 0.82$; and (F) $t>1.6\ {\rm
days}$: the light curve decays with an index of $2.4\pm0.4$
(see Fig.~\ref{fig:all_the_data}).

\citet{wyf+05} have argued that the fast decay during stage (D)
represents the normal afterglow, and the flattening at stage (E) is
caused by energy injection.  Then the stage (F) would indicate a
return to the normal afterglow evolution.  Here, however, we have
presented a different interpretation for the stages (D) and (E).  The
flux at stage (D) is considered to be suppressed by the inverse
Compton interaction between electrons in the forward shock and \xray\
flare photons, while the flux in stage (E) is the normal flux without
external inverse Compton process. This is motivated by the argument of
\citet{wlm06} suggesting that the \xray\ flare photons can interact
with the electrons in the forward shock regions via inverse Compton
scattering. The origin of the late-time \xray\ flares is unknown,
although a widely-held view is that they are due to the internal
shocks from late time engine activity. Since these \xray\ photons
would be coming, in this view, from a region different from (and
behind) the forward shock, we call this Inverse Compton (IC) process
an ``external IC process.''  In this case, the external IC process
will contribute significantly to the cooling of the forward shock
electrons, since the flare luminosity is much larger than the forward
shock (afterglow) luminosity, and the Compton parameter is determined
by the ratio of those two luminosities.  If the total radiated energy
at a given time is constant, when the energy radiated via the inverse
Compton process increases, the synchrotron radiation should
decrease. In effect, this can be viewed as the synchrotron radiation
having been suppressed by the inverse Compton process. Since the
$J$-band luminosity at this time is dominated by synchrotron
radiation, the observed flux would become smaller in the presence of
strong external IC processes. At the end of the flare, the external
inverse Compton suppression disappears, and the synchrotron radiation
can then return to its normal course.

If we assume that the averaged luminosity ratio between the flare and
the forward shock in the \xray\ is $k=L_{\rm IC,fl}/L_{\rm syn,f}$
(``fl'' is the flare and ``f'' denotes the forward shock), following
the similar definition for the Compton parameter which is the ratio of
the IC luminosity to the synchrotron luminosity $Y_{\rm SSC}=L_{\rm
IC,f}/L_{\rm syn,f}=(\eta \epsilon_e/\epsilon_B)^{1/2}$ where the
subscript ``SSC'' indicates the self-Compton scattering process
\citep{se+01}, we can get the new Compton parameter, considering the
external IC process, as $Y_{\rm IC,fl}=L_{\rm IC,total}/L_{\rm
syn,f}=[(L_{\rm IC,fl}+L_{\rm IC,f})/L_{\rm syn,f}]=[(k+1)\eta
\epsilon_e/\epsilon_B]^{1/2}$.  We can see an additional factor of
$(k+1)^{1/2}$ contributes to the Compton parameter for the external IC
process compared with the Compton parameter for the usual SSC
case. Because normally the parameter $k\gg 1$, we expect $Y_{\rm
IC,ext} > Y_{\rm SSC}$.  For the fast-cooling case, $\eta=1$, so we
have $Y_{\rm IC,fl,fast}=[(k+1)\epsilon_e/\epsilon_B]^{1/2}$ and
$Y_{\rm SSC,fast}=(\epsilon_e/\epsilon_B)^{1/2}$. The IC process will
affect the synchrotron radiation, and change the cooling frequency
$\nu_c$ for synchrotron radiation and the observed flux
$F_{\nu}\propto \nu_c^{1/2}$.  For synchrotron radiation, $\nu_c
\propto (1+Y)^{-2}$ and the external IC process will lead to a much
lower cooling frequency $\nu_c$ due to an increase in the value of the
Compton parameter. During the time period of stage (D) in Fig. 1, the
electrons are in the fast-cooling regime, $\nu_c < \nu_m < \nu_o$,
where $\nu_o$ is the observing frequency, so that the ratio of the
flux without external IC to the flux with external IC is
$\hat{S}={F_{\rm no}/F_{\rm IC}}=(1+Y_{\rm IC,fl,fast})/(1+Y_{\rm
SSC,fast})\approx (k+1)^{1/2}$, where $F_{\rm IC}$ is the flux with
external IC process considered and the flux without IC is $F_{\rm
no}=(1+Y_{\rm IC,ext})/(1+Y_{\rm SSC})F_{\rm IC,ext} \sim (k+1)^{1/2}
F_{\rm IC}$. Therefore, the observed flux in the optical/NIR band
during section (D) should be multiplied by a factor $[(1+k)^{1/2}]$ to
recover the flux that would be observed without IC suppression effects.

We can estimate the required luminosity ratio from the observed
suppression factor. Taking the power-law index for the electron energy
distribution to be $p=2.2$, the theoretically expected temporal index
is $\alpha=(3p-2)/4=1.15$ for $\nu_{o}>(\nu_c, \nu_m)$, the expected
regime for the optical afterglow during stage (E). Then we can
extrapolate the optical lightcurve back from $t=1.92\times 10^5$\,s
(where the flux is $f=9.14 \pm 1.75$\uJy) to the observer time
$t=2.7\times 10^4$\,s, where the observed flux is $f\approx 55$\,\uJy.
This theoretical flux, in the absence of the external IC process, will
be $9.14\cdot (1.92\times 10^5 /2.7\times 10^4)^{1.15} \simeq
81.0$\,\uJy.  Thus, the flux will be suppressed by a factor of
$\hat{S}\sim 1.6 \approx (1+k)^{1/2}$. From the observational point of
view, we invert this problem and solve for $k$, deriving $k\approx
1.6$.

Apparently, compared to the mean observed luminosity ratio between
the flares and the forward shock, $F_{\rm fl}/F_{\rm fs}\simg 10$
(see Fig.~\ref{fig:all_the_data}), this indicates within our picture
that only a small fraction of the flare photons have interacted with
the electrons in the afterglow.  A possible explanation could be
that the anisotropic distribution of the incoming flare photons in
the comoving frame of the afterglow shock \citep{wlm06}.  This
resultsing more head-on scattering which reduces the IC interaction.
Therefore, the suppression is relatively small and the optical flux
is only affected to a reduce degree.


\subsection{Radio Light-curve}
\label{sec:radio_light}

GRB\,050904 shows several similarities with GRB\,990123, including a
large isotropic-equivalent gamma-ray energy and a very bright optical
flash. In GRB\,990123, the radio emission was observed to peak at
$t=1$~day in the observer frame, and this emission was interpreted as
the radio emission from the reverse shock \citep{sp99a}.  Given the
important role of the reverse shock in our model (B) of GRB 050904, it
is interesting to consider its radio emission. However, our estimates
indicate that the reverse shock radio emission at early time would be
suppressed by the large circumburst density in our model, so that we
would not expect to observe a bright radio flare. In
Fig.~\ref{fig:light_curve_flare}, we have plotted the reverse shock
radio emission as a dotted blue line starting from $t=470$\,s, while
the solid blue line shows the combined radio emission from both the
forward shock and the reverse shock; as can be seen, the emission from
the reverse shock is negligible.

In Fig.~\ref{fig:light_curve_flare}, there is one averaged radio data
point centered at $t\sim2\times 10^5 $ seconds and $f_{\nu}\sim 30\
{\rm \mu Jy}$ (indicated with dashed lines ) based on the average of
multiple VLA observations spanning $\sim 6$ days \citep[see Fig. 2
of][]{fckn06}. Given the large dynamic range in time, we consider this
data point as an upper limit of the radio emission during that
period. We can see that our best-fit radio light curve for the forward
shock in both models (A) and (B) fits accommodates this upper limit
reasonably well, along with the other data points.


\subsection{Density and Energy Constraints}
\label{sec:energy_density}

The density obtained in our fit, $n\approx 84.4$~\percmcube\ for
model (A) and  $n\approx 212.4$~\percmcube\ for model (B), is
smaller than the $n=680$~\percmcube\ density derived by
\citet{fckn06}. Investigating our model fits, we find that the
discrepancy arises because their model peaks later in the radio than
ours.  Referring to the posterior distribution for the density
parameter in Figs.~\ref{fig:for_dis} and \ref{fig:for_rev_dis}, we
find that the most likely range (90\%-confidence) for density is
from 26 to 273.4~\percmcube\ for model (A) and from 87.8 to
270.6~\percmcube\ for model (B).  On the other hand, the light
curves for different densities demonstrate that there is no big
noticeable affect of these density changes on the \xray\ and
optical/NIR lightcurves.

The radio observations thus provide the best frequency for density
constraints, and fail to provide a tight constraint mainly because of
the lack of data (measurements) late times.

With regards to the circumburst density, we note that \citet{kka+05}
observed the spectrum and detected in it fine structure lines
including $\rm SiII^{*}$, which were taken to imply an electron
density of up to $10^{2.3\pm0.7}\ {\rm cm^{-3}}$.  The density
obtained from the spectral lines would thus be consistent with our
best-fit value for the density, and suggest that the observed lines
formed in region similar to that hosting the GRB itself.  However, we
note that several recent papers \citep{bp+05,pcb+06} have suggested
that these fine-structure transitions in GRB afterglows are excited by
radiative processes, rather than collisions, which would make the
density constraint irrelevant.

An important check on our model results, pointed out by
\citet{fckn06}, is to estimate the \xray\ luminosity from the \xray\
light curve at some fiducial time (usually at $t=10$ hours). At
$t=3.36$ days, the observed flux over the XRT energy range is $\sim
2.1\times 10^{-14}\ {\rm erg\ s^{-1}\ cm^{-2}}$. Calculating back to
$t\sim 10$ hours for the afterglow flux from synchrotron radiation,
the predicted flux is $f\sim 2.8\times 10^{-13}\ {\rm erg\ s^{-1}\
cm^{-2}}$. Therefore the \xray\ luminosity should be $L_{X,{\rm
iso}}=4\pi D_{L}^2 f (1+z)^{-\alpha+\beta-1} \sim 1.66\times 10^{46}\
{\rm erg\ s^{-1}}$, and the geometrically-corrected \xray\ luminosity
we find is different from and significantly lower than theirs. The
reason for this is that when they calculate the isotropic-equivalent energy, they
appear not to have included the $k$-correction factor
$(1+z)^{-\alpha+\beta-1}$. \citet{bkf03} also appear not to have
included the $k$-correction factor when doing statistics on the
isotropic-equivalent and geometrically-corrected \xray\ luminosity. Here we do
the statistics again after putting the correction, and find that the
geometrically-corrected \xray\ luminosities are still clustered, but
the peak value has shifted down to $L_{X,p}\sim 10^{44} {\rm erg/s}$,
reduced by a factor of five.

Once we have obtained the \xray\ luminosity, we can estimate the
kinetic energy with the best fitting parameters and the observed
spectral index. In model (A), we have best fitting parameters
$\epsilon_e=0.0309$, $\epsilon_{B,f}=0.198$ , $\beta \sim 1$,
$p=2.152$, and $\alpha=(3p-2)/4$. From the equation
\ref{eqn:kinetic_energy} in the appendix, we have $E_{52} \simeq 24.4
$, which is consistent with the kinetic energy derived from the fit,
$E_{K} = 22.4$. This is not surprising since the formula in the
appendix is only a shortcut to obtain the kinetic energy. In model
(B), we have the best fitting parameters $\epsilon_e=8.4\times
10^{-3}$, $\epsilon_{B,f}=5.7\times 10^{-3}$ , $\beta \sim 1$,
$p=2.243$, and $\alpha=(3p-2)/4$, and similarly we obtain $E_{52}
\simeq 148$. After considering the $k$-correction factor for the
\xray\ luminosity by \citet{fckn06}, we can re-estimate the kinetic
energy as $E_{52} \simeq 230$ with the parameters $\epsilon_e=0.02$,
and $\epsilon_B=0.015$, which is 3 times larger than their best-fit
kinetic energy $E_{52} \simeq 88$.

Considering our estimated opening half-angles of $\theta\simeq 0.13$
for both models (A) and (B), we can
calculate the geometrically-corrected \xray\ luminosity for
GRB\,050904 to be $L_{X}=L_{X,{\rm iso}}\times \theta^2/2
\approx 1.40\times 10^{44}\ {\rm erg\ s^{-1}}$, which falls
within the corrected luminosity range of low-redshift GRBs
\citep{bkf03}.

We now calculate the geometrically-corrected kinetic energy. Our
best-fit opening half-angle is $\theta\simeq 0.13$ for both models,
and the fitted kinetic energy for model (A) and (B) is $2.24\times
10^{53}$ ergs and $1.47\times 10^{54}$ ergs, so the
geometrically-corrected kinetic energy will be $1.84\times 10^{51}$
ergs and $1.17\times 10^{52}$ ergs for models (A) and (B),
respectively. Broadband modeling of 10 low-redshift bursts indicated
that the geometrically-corrected kinetic energies of two were
anomalously high, $2\times 10^{51}$ ergs, approximately 10 times
higher than for the other eight GRBs \citep{pk+02}. Our
geometrically-corrected kinetic energy for model (A) is comparable to
those anomalously-large kinetic energies seen from low-redshift
GRBs. The kinetic energy of model (B), on the other hand, is
significantly larger even than this.  Both models yield a relatively
large kinetic energy has been obtained for GRB\,050904. GRB\,050904
thus suggests that bursts at high redshift are somehow able to tap
into a higher-energy reservoir than the low-redshift events.


\subsection{Burst Energetics and Efficiency}
\label{sec:burst_eneregy}

The radiated isotropic-equivalent gamma-ray energy for GRB\,050904 is
$6.6\times 10^{53} < E_{\gamma,{\rm iso}} < 3.2 \times 10^{54}$ ergs
\citep{cmc+06}. If the isotropic-equivalent kinetic energy for the
afterglow of GRB\,050904, as we concluded in case (A), is $2.24\times
10^{53}$ ergs, we can estimate the GRB efficiency as $\zeta=E_{\gamma,
{\rm iso},52}/(E_{\gamma,{\rm iso},52}+E_{52})$ \citep{lz04}, which is
$74.7\%< \zeta < 93.5\%$ for GRB\,050904. If the isotropic-equivalent
kinetic energy of the afterglow, as in case (B), is $1.47\times
10^{54}$ ergs, then the corresponding GRB efficiency is roughly
$31.0\% < \zeta < 68.5\%$. In either case, this indicates that
GRB\,050904 has a high efficiency; however, such high efficiencies are
not unique.  \citet{lz04} found that a substantial number of GRBs have
high efficiency. For some bursts like GRB\,990705, the inferred
efficiency even reaches up to $99\%$. \citet{fp06} have shown that the
inferred efficiency can be reduced when inverse Compton effects are
taken into account \citep[see also][]{gkp+06}. Even so, there are
still several bursts which have a high observed efficiency, for
example, $94\%$ for GRB 050315 and $80\%$ for GRB 050416
\citep{zlpg+06}.


\section{Discussion}
\label{sec:disc}

The extreme interest in GRB\,050904 has motivated several groups to
analyze the burst data and suggest interpretations.  These works fall
into two categories: (1) Comparing the properties of GRB\,050904 with
other bursts; and (2) Making fits to the GRB\,050904 data either in
the framework of internal shock or external shock models. We give a
brief description below of the work of other groups and mention key
differences between their work and ours.

In category (1), \citet{kmk+06} made a composite $J$-band light curve
starting from $2\times 10^{-3}$ days to $\sim 23$ days. After applying
extinction correction, they shifted GRB\,050904 as well as
lower-redshift bursts to $z=1$, and made a comparison of afterglow
lightcurves. They found that GRB\,050904 is much brighter than other
GRBs at early times, but of roughly equal brightness at late
times. Thus they conclude that GRB\,050904 most likely is still a
normal GRB.

The other analyses are all in category (2). \citet{zdx+06} argued that
GRB\,050904 is a burst with extremely long central engine
activity. They put all the observed data within the framework of the
internal shock model. By contrast, we only treat the first several
hundred seconds (BAT) as  internal shock activity in our model
(corresponding to the stages (A) and (B) for model (A), and stage (A)
for model (B) in Fig. 1). The late-time \xray\ flares between $6\times
10^3$ and $6\times 10^4$ seconds may be due to internal shock, but we
have interpreted portions of this \xray\ emission as being due to the
forward-shock afterglow.

\citet{wyf+05} argued that the $t\approx 470$\,s flare is from
internal shocks on the basis of its fast decay (the temporal index is
$\alpha \approx 8.8$ relative to the trigger time) and also because
the optical-to-\xray\ emission of the flare cannot be described by a
synchrotron radiation model \citep{boer+05}. They made fits to all the
available $J$-band data. They argue that the slow-decay portion of the
lightcurve is due to energy injection. In our model, besides fitting
over all the additional bands (\xray\ and radio), we propose a new
mechanism for the flattening, namely that it is caused by the
suppression of the synchrotron radiation by the interaction between
the \xray\ flare photons and afterglow electrons. Separately, in our
model (B), we introduced a new reference time $t_{0}^{'}$ which
flattens the decay index, and allows us to interpret the $t\approx
470$\,s optical/\xray\ flare as arising in the reverse shock.

\citet{fckn06} made broadband model fits including the \xray,
NIR/optical, and radio data. The difference with our work is partly
that we have used the larger data set that later became available.
We have included two \xray\ data points as early as $t\sim 10^3$
seconds, and we also included the $H$ band data observed $t\sim 23$
days after the burst \citep{bcc+06}.  Since most of the other data
are concentrated around $t\sim 10^5$ seconds, the introduction of
these additional \xray\ and NIR data have some impact on the final
fitting result. The other major difference is that we have freed all
the possible parameters and applied the MCMC method for the global
fitting, making an efficient exploration of the full parameter
space, and providing the posterior distributions (including
confidence intervals) for each parameter.

\citet{ggck+06} argued that the power-law-like decay right after the
flare $t\approx 470$\,s should be interpreted as forward shock
emission. Because the extrapolation of the late time \xray\ flux to
early times is lower than the observed value, they found that a
wind-type environment was favored by the closure relationship for this
early-time segment (it should also be noted that their spectral index
is smaller than \citet{cmc+06}).  They propose a density-jump model
for the afterglow evolution: before a certain radius, the density goes
as $n \propto r^{-2}$ and after that, the density is a constant ISM
model. At the transition point a termination shock is formed which
lies around $R_t \sim 1.8 \times 10^{-2}$\,pc from the central engine.
In our model, we consider the same segment of data, but interpret it
differently. We argue that the flux between 600 and 800 seconds arises
from the combination of three sources: high-latitude emission of the
prompt emission, flux from the flare, and forward shock
emission. Reviewing the data closely, we see that actually there are
two other small flares between 800 and 2000 seconds, so the data
around $t \sim 1500$ seconds has the contribution from the flares and
the forward shock (see Fig. 2).  Once we subtract the flare
contribution from the observed data, we argue, the forward shock
contribution is what remains.  And in fact, we find that an
extrapolation of the late-time flux to early times is consistent with
this flare-subtracted early-time flux.

There remain some substantial differences in parameter values between
the two models that we have presented.  For example, the most likely
value for $\epsilon_{B,f}$ in model (A) is roughly $\sim 0.1$, but the
most likely value for $\epsilon_{B,f}$ in model (B) is around $5\times
10^{-3}$. It turns out that the main difference lies in the optical
flux at early times.  We take as an example the $J$-band light curve
for the best fitting parameter set in model (A). The flux slowly
decays with an index of $\alpha\approx 0.5$ before $t \sim 2\times
10^3$ and then follows a faster decay with a temporal index of
$\alpha\approx 1.1$ after $t \sim 2\times 10^3$. The break at $t \sim
2\times 10^3$ seconds is caused by the crossing of the electron's
typical frequency $\nu_m$ through the optical observing frequency.
Since we have $\nu_m \propto \epsilon_{B,f}^{1/2}$, then the smaller
$\epsilon_{B,f}$, the smaller the typical frequency, therefore the
earlier the crossing time. The fit in model (B) requires an earlier
break than in model (A), so a smaller value of $\epsilon_{B,f}$ is
expected in model (B). Since the observed flux from the forward shock
is the same in both models, we expect a higher kinetic energy in model
(B). Similarly, at the deceleration time, the reverse shock flux is
larger than that from the forward shock, so a larger $\epsilon_{B,r}$
is expected in the reverse shock.

We notice that in both our models we find a small value for
$\epsilon_e$, e.g., the best fitting $\epsilon_e=3.1\times 10^{-2}$
for model (A) and $\epsilon_e=8.4 \times 10^{-3}$ for model (B).
Considering the radiative correction factor $R=(t/t_0)^{17\epsilon_e
/48}$ (where $t_0$ is the deceleration time, and we set $t_0=300$
seconds for model (A) and $t_0=470$ seconds for model (B)), that is,
the amount by which the kinetic energy at time $t$ is reduced by
comparison to the kinetic energy at time $t_0$, we find $R=1.07$ for
model (A) and $R=1.02$ for model (B) at $t=1.0\times 10^5$ seconds
after the burst. Therefore, the radiative losses are mild in either
case, consistent with our assumption that the afterglow evolves in
an adiabatic fashion.

Our fitting results also show that the host galaxy dust extinction
is quite small, $A_V \sim 0.1$ mag or even smaller, consistent with
the results from \citet{kmk+06}.  They applied several different
dust models (MW, LMC and SMC) over the composite $J$-band, and all
the models suggested zero or negligible extinction. In the context
of GRB\,990123, it was suggested that the negligible extinction to
the burst was the result of dust destruction by the strong burst and
early-afterglow emission \citep{kkz+06}.


\section{CONCLUSIONS}
\label{sec:conclusions}

In this paper we have performed the most extensive multiband analysis
so far of the GRB\,050904 afterglow. We have considered two scenarios:
(A) Only forward shock emission is considered, and the flares peaking
at $t\approx470$\, after the burst are assumed to be due to internal shocks
(or are otherwise independent of the afterglow); and (B) The NIR and
\xray\ flares at $t\approx 470$\,s are ascribed to emission from the reverse
shock -- when the ejecta has swept up enough material and starts to
decelerate, the synchrotron radiation in the reverse shock produces
the optical flare, and the self-Compton scattering of synchrotron
photons generates the flare observed in the XRT and BAT energy
bands.

Combining the early afterglow data with late-time observations in the
\xray,\ optical and radio, and using a Markov chain Monte Carlo
method, we present a full characterization of the posterior
distributions (including confidence intervals) for the various
parameters of our model fits.  Our best-fit parameter values for model
(A) are $p=2.152$, $\epsilon_e=0.309$, $\epsilon_{B,f}=0.198$,
$n=84.4$~\percmcube, $\theta=0.128$, $E_{52}=22.4$, and
$A_V=0.0343$\,mag, with a reduced chi-squared value of $36.2/26
\approx 1.39$.  Our best-fit parameter values for model (B) are
$p=2.243$, $\epsilon_e=0.0084$, $\epsilon_{B,f}=5.7 \times 10^{-3}$,
$n=212.4$~\percmcube, $\theta=0.126$, $E_{52}=147$, $A_V=3.18
\times 10^{-2}$\,mag, $\Gamma_0=183.6$, and $\epsilon_{B,r}=0.50$,
with a reduced chi-squared value of $53.0/28 \approx 1.89$.  Note that
the subscripts $r$ and $f$ refer to the reverse shock and forward
shock respectively, $\epsilon_{B,f}=\epsilon_{B,r}/R_B^2$, and we have
assumed $\epsilon_{e,f}=\epsilon_{e,r}$.

We have compared the density, the geometrically-corrected kinetic
energy and the \xray\ luminosity at $t=10$~hours derived here for our
two models of GRB\,050904 against those values for other bursts, as
derived from afterglow modeling.  The results for both models show
that although the \xray\ luminosity of GRB\,050904 falls within the
range for low-redshift GRBs, the density and geometrically-corrected
kinetic energy are both above the typical values for low-redshift
GRBs, which suggests that GRB\,050904 may be a member of a distinct
population of high-redshift, higher kinetic-energy bursts, whose
properties differ from those of low-redshift GRBs. A clear preference
between our (A) and (B) models is hard to establish at present, since
there is only one high-redshift GRB known.  One would like access to
several more high-redshift GRB observational datasets before
attempting to discriminate between the two models.

It is estimated that $\sim 7\%-40\%$ GRBs are located at $z>5$
\citep{jlfp+05}, and detection rate simulations by \citet{gmaz+04}
indicate that \swift\ could detect GRBs out to redshift $z\sim 30$, if
they are present. \citet{bl+06} also predict that $10\%$ of the
\swift\ GRBs originate at $z>5$. It appears that one can realistically
expect a handful (5 to 10) of additional high-redshift GRB detections
with rapid follow-up in the next few years of the \swift\ mission.  In
this case, the consistent application of MCMC methods, as used here,
will lead efficiently to a set of statistically well-quantified,
posterior parameter distributions and confidence intervals. This would
enable a statistically meaningful comparison of high-redshift and
low-redshift GRB parameters, which might well lead us to a definite
understanding of the physics and environments of GRBs as a function of
redshift, up to the highest redshifts detected.  This would also have
a substantial impact on the study of the large scale structure and
star formation processes throughout the Universe, and the
properties of the cosmic reionization at $z\sim 6$.


\acknowledgements

L.~J. Gou thanks G. Cusumano for providing the \xray\ data for this
burst, as well as J. Cummings, B. Zhang, S. Kobayashi, Zh. Li,
X.Y. Wang, and J.F. Wang for the helpful discussion. L.~J. Gou also
thanks the support of Sigma-Xi Fellowship. This research has been
supported in part through NASA NAG5-13286. This research has used the
resources of High Performance Computing (HPC) Group at the Penn State
University.


\begin{thebibliography}{71}
\expandafter\ifx\csname natexlab\endcsname\relax\def\natexlab#1{#1}\fi

\bibitem[{{Akerlof} {et~al.}(1999){Akerlof}, {Balsano}, {Barthelemy}, {Bloch},
  {Butterworth}, {Casperson}, {Cline}, {Fletcher}, {Frontera}, {Gisler},
  {Heise}, {Hills}, {Kehoe}, {Lee}, {Marshall}, {McKay}, {Miller}, {Piro},
  {Priedhorsky}, {Szymanski}, \& {Wren}}]{abb+99}
{Akerlof}, C. {et al.}\  1999, \nat, 398, 400

\bibitem[{{Berger}(2007)}]{be+07}
{Berger}, E. 2007, GRB Coordinates Network, 6018, 1

\bibitem[{{Berger} {et~al.}(2006){Berger}, {Chary}, {Cowie}, {Price},
  {Schmidt}, {Fox}, {Cenko}, {Djorgovski}, {Soderberg}, {Kulkarni}, {McCarthy},
  {Gladders}, {Peterson}, \& {Barger}}]{bcc+06}
{Berger}, E. {et al.}\  2006, ArXiv Astrophysics e-prints, astro-ph/0603689

\bibitem[{{Berger} {et~al.}(2003){Berger}, {Kulkarni}, \& {Frail}}]{bkf03}
{Berger}, E., {Kulkarni}, S.~R., \& {Frail}, D.~A. 2003, \apj, 590, 379

\bibitem[{{Berger} {et~al.}(2005){Berger}, {Penprase}, {Fox}, {Kulkarni},
  {Hill}, {Schaefer}, \& {Reed}}]{bp+05}
{Berger}, E. {et al.}\  2005, ArXiv Astrophysics e-prints, astro-ph/0512280

\bibitem[{{Blandford} \& {McKee}(1976)}]{bm+76}
{Blandford}, R.~D. \& {McKee}, C.~F. 1976, Physics of Fluids, 19, 1130

\bibitem[{{Bo{\"e}r} {et~al.}(2006){Bo{\"e}r}, {Atteia}, {Damerdji}, {Gendre},
  {Klotz}, \& {Stratta}}]{boer+05}
{Bo{\"e}r}, M. {et al.}\  2006, \apjl, 638, L71

\bibitem[{{Bromm} \& {Loeb}(2006)}]{bl+06}
{Bromm}, V. \& {Loeb}, A. 2006, \apj, 642, 382

\bibitem[{{Burrows} \& {Racusin}(2007)}]{br+07}
{Burrows}, D.~N. \& {Racusin}, J. 2007, ArXiv Astrophysics e-prints,
  astro-ph/0702633

\bibitem[{{Cenko} {et~al.}(2007){Cenko}, {Soderberg}, {Frail}, \&
  {Fox}}]{csff+07}
{Cenko}, S.~B. {et al.}\  2007, GRB Coordinates Network, 6186, 1

\bibitem[{{Chincarini} {et~al.}(2007){Chincarini}, {Moretti}, {Romano},
  {Falcone}, {Morris}, {Racusin}, {Campana}, {Guidorzi}, {Tagliaferri},
  {Burrows}, {Pagani}, {Stroh}, {Grupe}, {Capalbi}, {Cusumano}, {Gehrels},
  {Giommi}, {La Parola}, {Mangano}, {Mineo}, {Nousek}, {O'Brien}, {Page},
  {Perri}, {Troja}, {Willingale}, \& {Zhang}}]{cmrf+07}
{Chincarini}, G. {et al.}\  2007, ArXiv Astrophysics e-prints, astro-ph/0702371

\bibitem[{{Cohen} {et~al.}(1998){Cohen}, {Piran}, \& {Sari}}]{cps+98}
{Cohen}, E., {Piran}, T., \& {Sari}, R. 1998, \apj, 509, 717

\bibitem[{{Cusumano} {et~al.}(2006){Cusumano}, {Mangano}, {Chincarini},
  {Panaitescu}, {Burows}, {La Parola}, {Sakamoto}, {Campana}, {Mineo},
  {Tagliaferri}, {Angelini}, {Barthelemy}, {Beardmore}, {Boyd}, {Cominsky},
  {Gronwall}, {Fenimore}, {Gehrels}, {Giommi}, {Goad}, {Hurley}, {Kennea},
  {Mason}, {Marshall}, {M\'esz\'aros}, {Nousek}, {Osborne}, {Palmer}, {Roming},
  {Wells}, {White}, \& {Zhang}}]{cmc+06}
{Cusumano}, G. {et al.}\  2006, \nat, 440, 164

\bibitem[{{Cusumano} {et~al.}(2007){Cusumano}, {Mangano}, {Chincarini},
  {Panaitescu}, {Burrows}, {La Parola}, {Sakamoto}, {Campana}, {Mineo},
  {Tagliaferri}, {Angelini}, {Barthelemy}, {Beardmore}, {Boyd}, {Cominsky},
  {Gronwall}, {Fenimore}, {Gehrels}, {Giommi}, {Goad}, {Hurley}, {Immler},
  {Kennea}, {Mason}, {Marshall}, {Meszaros}, {Nousek}, {Osborne}, {Palmer},
  {Roming}, {Wells}, {White}, \& {Zhang}}]{cmc+06b}
--- 2007, \aap, 462, 73

\bibitem[{{Dai} {et~al.}(2007){Dai}, {Halpern}, {Morgan}, {Armstrong},
  {Mirabal}, {Haislip}, {Reichart}, \& {Stanek}}]{dhma+07}
{Dai}, X. {et al.}\  2007, \apj, 658, 509

\bibitem[{{Falcone} {et~al.}(2006){Falcone}, {Burrows}, {Lazzati}, {Campana},
  {Kobayashi}, {Zhang}, {M{\'e}sz{\'a}ros}, {Page}, {Kennea}, {Romano},
  {Pagani}, {Angelini}, {Beardmore}, {Capalbi}, {Chincarini}, {Cusumano},
  {Giommi}, {Goad}, {Godet}, {Grupe}, {Hill}, {La Parola}, {Mangano},
  {Moretti}, {Nousek}, {O'Brien}, {Osborne}, {Perri}, {Tagliaferri}, {Wells},
  \& {Gehrels}}]{fblc+06}
{Falcone}, A.~D. {et al.}\  2006, \apj, 641, 1010

\bibitem[{{Fan} \& {Piran}(2006)}]{fp06}
{Fan}, Y. \& {Piran}, T. 2006, \mnras, 369, 197

\bibitem[{{Fan} {et~al.}(2002){Fan}, {Dai}, {Huang}, \& {Lu}}]{fdhl+02}
{Fan}, Y.-Z. {et al.}\  2002, Chinese Journal of Astronony and Astrophysics, 2,
  449

\bibitem[{{Frail} {et~al.}(2006){Frail}, {Cameron}, {Kasliwal}, {Nakar},
  {Price}, {Berger}, {Gal-Yam}, {Kulkarni}, {Fox}, {Soderberg}, {Schmidt},
  {Ofek}, \& {Cenko}}]{fckn06}
{Frail}, D.~A. {et al.}\  2006, \apjl, 646, L99

\bibitem[{{Frail} {et~al.}(2000){Frail}, {Waxman}, \& {Kulkarni}}]{fwk+00}
{Frail}, D.~A., {Waxman}, E., \& {Kulkarni}, S.~R. 2000, \apj, 537, 191

\bibitem[{{Gehrels} {et~al.}(2004){Gehrels}, {Chincarini}, {Giommi}, {Mason},
  {Nousek}, {Wells}, {White}, {Barthelmy}, {Burrows}, {Cominsky}, {Hurley},
  {Marshall}, {M{\'e}sz{\'a}ros}, {Roming}, {Angelini}, {Barbier}, {Belloni},
  {Campana}, {Caraveo}, {Chester}, {Citterio}, {Cline}, {Cropper}, {Cummings},
  {Dean}, {Feigelson}, {Fenimore}, {Frail}, {Fruchter}, {Garmire}, {Gendreau},
  {Ghisellini}, {Greiner}, {Hill}, {Hunsberger}, {Krimm}, {Kulkarni}, {Kumar},
  {Lebrun}, {Lloyd-Ronning}, {Markwardt}, {Mattson}, {Mushotzky}, {Norris},
  {Osborne}, {Paczynski}, {Palmer}, {Park}, {Parsons}, {Paul}, {Rees},
  {Reynolds}, {Rhoads}, {Sasseen}, {Schaefer}, {Short}, {Smale}, {Smith},
  {Stella}, {Tagliaferri}, {Takahashi}, {Tashiro}, {Townsley}, {Tueller},
  {Turner}, {Vietri}, {Voges}, {Ward}, {Willingale}, {Zerbi}, \&
  {Zhang}}]{gcg+04}
{Gehrels}, N. {et al.}\  2004, \apj, 611, 1005

\bibitem[{{Gelman} \& {Rubin}(1992)}]{gr+92}
{Gelman}, A. \& {Rubin}, D. 1992, Stat. Sci., 7, 457

\bibitem[{{Gendre} {et~al.}(2007){Gendre}, {Galli}, {Corsi}, {Klotz}, {Piro},
  {Stratta}, {Boer}, \& {Damerdji}}]{ggck+06}
{Gendre}, B. {et al.}\  2007, \aap, 462, 565

\bibitem[{{Ghirlanda} {et~al.}(2004){Ghirlanda}, {Ghisellini}, \&
  {Lazzati}}]{ggl+04}
{Ghirlanda}, G., {Ghisellini}, G., \& {Lazzati}, D. 2004, \apj, 616, 331

\bibitem[{{Gou} {et~al.}(2001){Gou}, {Dai}, {Huang}, \& {Lu}}]{gdhl+01}
{Gou}, L.~J. {et al.}\  2001, \aap, 368, 464

\bibitem[{{Gou} {et~al.}(2004){Gou}, {M{\'e}sz{\'a}ros}, {Abel}, \&
  {Zhang}}]{gmaz+04}
--- 2004, \apj, 604, 508

\bibitem[{{Granot} {et~al.}(2006){Granot}, {K{\"o}nigl}, \& {Piran}}]{gkp+06}
{Granot}, J., {K{\"o}nigl}, A., \& {Piran}, T. 2006, \mnras, 370, 1946

\bibitem[{{Granot} \& {Sari}(2002)}]{gs+02}
{Granot}, J. \& {Sari}, R. 2002, \apj, 568, 820

\bibitem[{{Haislip} {et~al.}(2006){Haislip}, {Nysewander}, {Reichart}, {Levan},
  {Tanvir}, {Cenko}, {Fox}, {Price}, {Castro-Tirado}, {Gorosabel}, {Evans},
  {Figueredo}, {MacLeod}, {Kirschbrown}, {Jelinek}, {Guziy}, {de Ugarte
  Postigo}, {Cypriano}, {LaCluyze}, {Graham}, {Priddey}, {Chapman}, {Rhoads},
  {Fruchter}, {Lamb}, {Kouveliotou}, {Wijers}, {Schmidt}, {Soderberg},
  {Kulkarni}, {Harrison}, {Moon}, {Gal-Yam}, {Kasliwal}, {Hudec}, {Vitek},
  {Kubanek}, {Crain}, {Foster}, {Bayliss}, {Clemens}, \& {Bartelme}}]{hnr+06}
{Haislip}, J. {et al.}\  2006, \nat, 440, 181

\bibitem[{{Haislip} {et~al.}(2005){Haislip}, {Reichart}, {Cypriano}, {Pizzaro},
  {Lacluyze}, {Rhoads}, \& {Figueredo}}]{gcn.3914}
--- 2005, GRB Coordinates Network, 3914, 1

\bibitem[{{Harrison} {et~al.}(1999){Harrison}, {Bloom}, {Frail}, {Sari},
  {Kulkarni}, {Djorgovski}, {Axelrod}, {Mould}, {Schmidt}, {Wieringa}, {Wark},
  {Subrahmanyan}, {McConnell}, {McCarthy}, {Schaefer}, {McMahon}, {Markze},
  {Firth}, {Soffitta}, \& {Amati}}]{hbfs+99}
{Harrison}, F.~A. {et al.}\  1999, \apjl, 523, L121

\bibitem[{{Jakobsson} {et~al.}(2006){Jakobsson}, {Levan}, {Fynbo}, {Priddey},
  {Hjorth}, {Tanvir}, {Watson}, {Jensen}, {Sollerman}, {Natarajan},
  {Gorosabel}, {Castro Cer{\'o}n}, {Pedersen}, {Pursimo}, {{\'A}rnad{\'o}ttir},
  {Castro-Tirado}, {Davis}, {Deeg}, {Fiuza}, {Mykolaitis}, \&
  {Sousa}}]{jlfp+05}
{Jakobsson}, P. {et al.}\  2006, \aap, 447, 897

\bibitem[{{Kann} {et~al.}(2006){Kann}, {Klose}, \& {Zeh}}]{kkz+06}
{Kann}, D.~A., {Klose}, S., \& {Zeh}, A. 2006, \apj, 641, 993

\bibitem[{{Kann} {et~al.}(2007){Kann}, {Masetti}, \& {Klose}}]{kmk+06}
{Kann}, D.~A., {Masetti}, N., \& {Klose}, S. 2007, \aj, 133, 1187

\bibitem[{{Kawai} {et~al.}(2006){Kawai}, {Kosugi}, {Aoki}, {Yamada}, {Totani},
  {Ohta}, {Iye}, {Hattori}, {Aoki}, {Furusawa}, {Hurley}, {Kawabata},
  {Kobayashi}, {Komiyama}, {Mizumoto}, {Nomoto}, {Noumaru}, {Ogasawara},
  {Sato}, {Sekiguchi}, {Shirasaki}, {Suzuki}, {Takata}, {Tamagawa}, {Terada},
  {Watanabe}, {Yatsu}, \& {Yoshida}}]{kka+05}
{Kawai}, N. {et al.}\  2006, \nat, 440, 184

\bibitem[{{Kobayashi} \& {Zhang}(2007)}]{kz+06}
{Kobayashi}, S. \& {Zhang}, B. 2007, \apj, 655, 973

\bibitem[{{Kobayashi} {et~al.}(2007){Kobayashi}, {Zhang}, {M{\'e}sz{\'a}ros},
  \& {Burrows}}]{kzmb+05}
{Kobayashi}, S. {et al.}\  2007, \apj, 655, 391

\bibitem[{{Kumar} \& {Panaitescu}(2000)}]{kp+00}
{Kumar}, P. \& {Panaitescu}, A. 2000, \apjl, 541, L9

\bibitem[{{Kumar} \& {Panaitescu}(2003)}]{kp+03}
--- 2003, \mnras, 346, 905

\bibitem[{{Lamb} \& {Reichart}(2000)}]{lr+00}
{Lamb}, D.~Q. \& {Reichart}, D.~E. 2000, \apj, 536, 1

\bibitem[{{Liang} {et~al.}(2006){Liang}, {Zhang}, {O'Brien}, {Willingale},
  {Angelini}, {Burrows}, {Campana}, {Chincarini}, {Falcone}, {Gehrels}, {Goad},
  {Grupe}, {Kobayashi}, {M{\'e}sz{\'a}ros}, {Nousek}, {Osborne}, {Page}, \&
  {Tagliaferri}}]{lzb+06}
{Liang}, E.~W. {et al.}\  2006, \apj, 646, 351

\bibitem[{{Lloyd-Ronning} \& {Zhang}(2004)}]{lz04}
{Lloyd-Ronning}, N.~M. \& {Zhang}, B. 2004, \apj, 613, 477

\bibitem[{{M{\'e}sz{\'a}ros}(2006)}]{mm+06}
{M{\'e}sz{\'a}ros}, P. 2006, Reports of Progress in Physics, 69, 2259

\bibitem[{{Nava} {et~al.}(2007){Nava}, {Ghisellini}, {Ghirlanda}, {Cabrera},
  {Firmani}, \& {Avila-Reese}}]{nggc+07}
{Nava}, L. {et al.}\  2007, ArXiv Astrophysics e-prints,astro-ph/0701705

\bibitem[{{Panaitescu}(2005)}]{p05}
{Panaitescu}, A. 2005, \mnras, 363, 1409

\bibitem[{{Panaitescu} \& {Kumar}(2001{\natexlab{a}})}]{pk+01}
{Panaitescu}, A. \& {Kumar}, P. 2001{\natexlab{a}}, \apjl, 560, L49

\bibitem[{{Panaitescu} \& {Kumar}(2001{\natexlab{b}})}]{pk+01b}
--- 2001{\natexlab{b}}, \apj, 554, 667

\bibitem[{{Panaitescu} \& {Kumar}(2002)}]{pk+02}
--- 2002, \apj, 571, 779

\bibitem[{{Panaitescu} \& {M\'esz\'aros}(1998)}]{pm+97}
{Panaitescu}, A. \& {M\'esz\'aros}, P. 1998, \apjl, 493, L31+

\bibitem[{{Prochaska} {et~al.}(2006){Prochaska}, {Chen}, \& {Bloom}}]{pcb+06}
{Prochaska}, J.~X., {Chen}, H.-W., \& {Bloom}, J.~S. 2006, \apj, 648, 95

\bibitem[{{Sari}(1997)}]{s+97}
{Sari}, R. 1997, \apjl, 489, L37+

\bibitem[{{Sari} \& {Esin}(2001)}]{se+01}
{Sari}, R. \& {Esin}, A.~A. 2001, \apj, 548, 787

\bibitem[{{Sari} \& {Piran}(1999)}]{sp99a}
{Sari}, R. \& {Piran}, T. 1999, \apjl, 517, L109

\bibitem[{{Sari} {et~al.}(1999){Sari}, {Piran}, \& {Halpern}}]{sph+99}
{Sari}, R., {Piran}, T., \& {Halpern}, J.~P. 1999, \apjl, 519, L17

\bibitem[{{Sari} {et~al.}(1998){Sari}, {Piran}, \& {Narayan}}]{spn+98}
{Sari}, R., {Piran}, T., \& {Narayan}, R. 1998, \apjl, 497, L17+

\bibitem[{{Sato} {et~al.}(2007){Sato}, {Yamazaki}, {Ioka}, {Sakamoto},
  {Takahashi}, {Nakazawa}, {Nakamura}, {Toma}, {Hullinger}, {Tashiro},
  {Parsons}, {Krimm}, {Barthelmy}, {Gehrels}, {Burrows}, {O'Brien}, {Osborne},
  {Chincarini}, \& {Lamb}}]{syis+07}
{Sato}, G. {et al.}\  2007, \apj, 657, 359

\bibitem[{{Tagliaferri} {et~al.}(2005){Tagliaferri}, {Antonelli}, {Chincarini},
  {Fern{\'a}ndez-Soto}, {Malesani}, {Della Valle}, {D'Avanzo}, {Grazian},
  {Testa}, {Campana}, {Covino}, {Fiore}, {Stella}, {Castro-Tirado},
  {Gorosabel}, {Burrows}, {Capalbi}, {Cusumano}, {Conciatore}, {D'Elia},
  {Filliatre}, {Fugazza}, {Gehrels}, {Goldoni}, {Guetta}, {Guziy}, {Held},
  {Hurley}, {Israel}, {Jel{\'{\i}}nek}, {Lazzati}, {L{\'o}pez-Echarri},
  {Melandri}, {Mirabel}, {Moles}, {Moretti}, {Mason}, {Nousek}, {Osborne},
  {Pellizza}, {Perna}, {Piranomonte}, {Piro}, {de Ugarte Postigo}, \&
  {Romano}}]{tac+05}
{Tagliaferri}, G. {et al.}\  2005, \aap, 443, L1

\bibitem[{{Totani} {et~al.}(2006){Totani}, {Kawai}, {Kosugi}, {Aoki}, {Yamada},
  {Iye}, {Ohta}, \& {Hattori}}]{tkk+06}
{Totani}, T. {et al.}\  2006, \pasj, 58, 485

\bibitem[{{Verde} {et~al.}(2003){Verde}, {Peiris}, {Spergel}, {Nolta},
  {Bennett}, {Halpern}, {Hinshaw}, {Jarosik}, {Kogut}, {Limon}, {Meyer},
  {Page}, {Tucker}, {Wollack}, \& {Wright}}]{vps+03}
{Verde}, L. {et al.}\  2003, \apjs, 148, 195

\bibitem[{{Wang} {et~al.}(2001){Wang}, {Dai}, \& {Lu}}]{wdl+01}
{Wang}, X.~Y., {Dai}, Z.~G., \& {Lu}, T. 2001, \apj, 556, 1010

\bibitem[{{Wang} {et~al.}(2006){Wang}, {Li}, \& {M{\'e}sz{\'a}ros}}]{wlm06}
{Wang}, X.-Y., {Li}, Z., \& {M{\'e}sz{\'a}ros}, P. 2006, \apjl, 641, L89

\bibitem[{{Waxman}(1997)}]{wax+97}
{Waxman}, E. 1997, \apjl, 491, L19+

\bibitem[{{Wei} {et~al.}(2006){Wei}, {Yan}, \& {Fan}}]{wyf+05}
{Wei}, D.~M., {Yan}, T., \& {Fan}, Y.~Z. 2006, \apjl, 636, L69

\bibitem[{{Wu} {et~al.}(2003){Wu}, {Dai}, {Huang}, \& {Lu}}]{wdhl+03}
{Wu}, X.~F. {et al.}\  2003, \mnras, 342, 1131

\bibitem[{{Wu} {et~al.}(2005){Wu}, {Dai}, {Huang}, \& {Lu}}]{wdhl+05}
--- 2005, \apj, 619, 968

\bibitem[{{Yost} {et~al.}(2003){Yost}, {Harrison}, {Sari}, \&
  {Frail}}]{yhsf+03}
{Yost}, S.~A. {et al.}\  2003, \apj, 597, 459

\bibitem[{{Zhang} {et~al.}(2006){Zhang}, {Fan}, {Dyks}, {Kobayashi},
  {M\'esz\'aros}, {Burrows}, {Nousek}, \& {Gehrels}}]{zfd+05}
{Zhang}, B. {et al.}\  2006, \apj, 642, 354

\bibitem[{{Zhang} {et~al.}(2003){Zhang}, {Kobayashi}, \&
  {M{\'e}sz{\'a}ros}}]{zkm+03}
{Zhang}, B., {Kobayashi}, S., \& {M{\'e}sz{\'a}ros}, P. 2003, \apj, 595, 950

\bibitem[{{Zhang} {et~al.}(2007){Zhang}, {Liang}, {Page}, {Grupe}, {Zhang},
  {Barthelmy}, {Burrows}, {Campana}, {Chincarini}, {Gehrels}, {Kobayashi},
  {M\'esz\'aros}, {Moretti}, {Nousek}, {O'Brien}, {Osborne}, {Roming},
  {Sakamoto}, {Schady}, \& {Willingale}}]{zlpg+06}
{Zhang}, B. {et al.}\  2007, \apj, 655, 989

\bibitem[{{Zou} {et~al.}(2006){Zou}, {Dai}, \& {Xu}}]{zdx+06}
{Zou}, Y.~C., {Dai}, Z.~G., \& {Xu}, D. 2006, \apj, 646, 1098

\bibitem[{{Zou} {et~al.}(2005){Zou}, {Wu}, \& {Dai}}]{zwd+05}
{Zou}, Y.~C., {Wu}, X.~F., \& {Dai}, Z.~G. 2005, \mnras, 363, 93

\end{thebibliography}


\appendix


\section{Self-Absorption Frequency}
\label{self_ab}

Once we consider radio emission in our afterglow models, the
self-absorption frequency becomes an important parameter to consider.
Below we give the expressions for the self-absorption frequency in the
different regimes.

If we assume that the electron distribution follows a power law
$N(\gamma_e)=N_{\gamma} \gamma_{e}^{-p}$ where $\gamma_1 < \gamma_e <
\gamma_2$, then the self-absorption coefficient for the various
possible regimes is \citep{wdhl+03}:
\begin{equation}
  \label{eq:A_k_nu}
  k_{\nu}=\frac{q_e}{B}N_\gamma \left\{
  \begin{array}{ll}
    c_1 \gamma_1^{-(p+4)} \left( \frac{\nu}{\nu_1} \right) ^{-5/3} & \nu \ll \nu_1 \\
    c_2 \gamma_1^{-(p+4)} \left( \frac{\nu}{\nu_1} \right) ^{-(p+4)/2} & \nu_1 \ll \nu \ll \nu_2, \\
    c_3 \gamma_2^{-(p+4)} \left( \frac{\nu}{\nu_2} \right) ^{-5/2} e ^{-\nu/\nu_2} & \nu \gg \nu_2
  \end{array}
  \right.
\end{equation}
where $c_1=\frac{32 \pi^2}{9\times 2^{1/3}\Gamma(1/3)}
\frac{p+2}{p+2/3}$, $c_2=\frac{2\sqrt{3}\pi}{9}2^{p/2}
(p+\frac{10}{3})\Gamma(\frac{3p+2}{12}) \Gamma(\frac{3p+10}{12})$,
$c_3=\frac{2\sqrt{6}\pi^{3/2}}{9}(p+2)$, $\nu_1$ and $\nu_2$ are the
typical synchrotron frequencies of electrons with the Lorentz factor
$\gamma_1$ and $\gamma_2$, respectively, and $\Gamma(x)$ is the Gamma
function.

Following the definition of $\kappa_{\nu}(\nu_a) l=1$ for the
self-absorption frequency (where $l=R/\Gamma$ is the thickness of the
shell), we can find the self-absorption for the forward shock
region. The self-absorption frequency for the reverse shock has a
similar form, with the difference that the quantities specific to the
forward shock region should be replaced with those specific to the
reverse shock.

Fast Cooling: $\nu_a < \nu_c < \nu_m$,
\begin{equation}
\nu_a^{(1)}=\nu_c \left( {c_1 q_e n_0 R \over 3 B \gamma_c^5
}\right)^{3/5} \\
\end{equation}
where the superscript ``$i$'' denotes the different regimes.

Fast Cooling: $\nu_c < \nu_a < \nu_m$,
\begin{eqnarray}
\nu_a^{(2)}&=&\nu_c \left( {c_2  q_e n_0 R \over 3 B \gamma_c^5
}\right)^{1/3} \\ \nu_a^{(2)} &=&\nu_c^{4/9} (\nu_a^{(1)})^{5/9}
\end{eqnarray}

Fast Cooling: $\nu_c < \nu_m < \nu_a$,
\begin{eqnarray}
\nu_a^{(3)}&=&\nu_m \left( {c_2 q_e n_0 R  \gamma_c \over 3 B
\gamma_m^6 }\right)^{2/(p+5)}\\ \nu_a^{(3)}&=&[(\nu_a^{(1)})^{10/3}
\nu_c^{8/3} \nu_m^{p-1} ]^{1/(p+5)}
\end{eqnarray}

Slow Cooling: $\nu_a < \nu_m < \nu_c$
\begin{equation}
\nu_a^{(4)}=\nu_m \left( {c_1  (p-1) q_e n_0 R \over 3 B \gamma_m^5
}\right)^{3/5} \\
\end{equation}
We notice that $\nu_a^{4}/ \nu_a^{1} = (p-1)^{3/5}
(\nu_c/\nu_m)^{1/2} $. To keep the continuity of the flux while the afterglow
transits from the fast-cooling regime to the slow-cooling regime, we
divide $\nu_a^{(4)}$ by a factor of $(p-1)^{3/5}$.

Slow Cooling: $\nu_m < \nu_a < \nu_c$,
\begin{eqnarray}
\nu_a^{(5)}=\nu_m \left( {c_2  (p-1) q_e n_0 R \over 3 B \gamma_m^5
}\right)^{2/(p+4)}\\ \nu_a^{(5)}= [(\nu_a^{(4)})^{10}
\nu_m^{(3p+2)}]^{1/(3(p+4))}
\end{eqnarray}

Slow Cooling: $\nu_m < \nu_c < \nu_a$,
\begin{eqnarray}
\nu_a^{(6)}=\nu_c \left( {c_2 (p-1) q_e  n_0 R  \gamma_m^{p-1} \over
3 B \gamma_c^{p+4} }\right)^{2/(p+5)} \\
\nu_a^{(6)}=[(\nu_a^{(1)})^{10/3} \nu_c^{8/3} \nu_m^{p-1}
]^{1/(p+5)}
\end{eqnarray}
it can be shown that the self-absorption frequency in the regime
$\nu_m < \nu_c < \nu_a$ has the same form as that for the
fast-cooling case, $\nu_c < \nu_m < \nu_a$.

Because the observer time, $t_{obs}$, at $z=0$ is connected to the
time in the source frame, $t_s$, at redshift $z$ by the relation
$t_{obs}=(1+z) t_{s}$, we have the redshift dependence for main
characteristic quantities: the shock radius $R \propto t_{s}^{1/4}
\propto (1+z)^{-1/4}$, the shock Lorentz factor $\Gamma \propto
t_s^{-3/8} \propto (1+z)^{3/8}$, the magnetic field $B \propto
\Gamma \propto (1+z)^{3/8}$, the typical Lorentz factor $\gamma_m
\propto \Gamma \propto (1+z)^{3/8}$, the cooling Lorentz factor
$\gamma_c \propto t_s^{-1} \Gamma^{-3} \propto (1+z)^{-1/8}$. In
addition, we have $\nu_m \propto (1+z)^{1/2}$ and $\nu_c \propto
(1+z)^{-1/2}$ (see Eqn. \ref{eqn:forward}). Substituting these
dependence into the relations for the self-absorptions above, the
redshift dependence for the self-absorptions is $\nu_a^{(1)} \propto
(1+z)^{-1/2}, \nu_a^{(2)} \propto (1+z)^{-1/2}, \nu_a^{3} \propto
(1+z)^{(p-7)/[2(p+5)]}, \nu_a^{(4)} \propto (1+z)^{-1}, \nu_a^{(5)}
\propto (1+z)^{(p-6)/[2(p+4)]}$,~and $\nu_a^{(6)}=\nu_a^{(3)}
\propto (1+z)^{(p-7)/[2(p+5)]}$.


\section{Inverse Compton Spectrum}
\label{app:ic}

As described by \citet{se+01}, the inverse Compton flux can be
calculated from the following double integral:
\begin{equation}
f_{\nu}^{IC} =R \sigma_T \int_{\gamma_m}^{\infty} d \g N(\g)
\int_{0}^{x_0} dx f_{\nu_s} (x) \\
\label{inverse_compton}
\end{equation}
where $N(\g)$ is the electron distribution in the shocked shell,
$f_{\nu_s}(x)$ is the seed photon flux, and $x$ is defined as $x=
\nu/4 \g^2 \nu_s$ where the subscript ``$s$'' denotes the seed photon.

\subsection{$\nu_c < \nu_a < \nu_m$}

The distribution of seed photons is described by the synchrotron
spectrum, a broken power law with the characteristic quantities (Sari
et al. 1998). Then the inner integral in Eqn.  (\ref{inverse_compton})
gives:
\begin{equation}
\label{int1} I = \left\{ \begin{array}{l@{\quad \quad}l}
I_1\simeq\frac{5}{3} f_{max} x_0 \left(\frac{\nu}{4 \g^2 \nu_a^{(2)} x_0}\right), &  \nu < 4 \g^2 \nu_a^{(2)} x_0 \\
I_2 \simeq \frac{2}{3} f_{max} x_0 \left(\frac{\nu}{4 \g^2 \nu_m x_0}\right)^{\frac{-1}{2}},& 4 \g^2 \nu_a^{(2)} x_0 < \nu < 4 \g^2 \nu_m x_0 \\
I_3 \simeq \frac{2}{(p+2)} f_{max} x_0 \left(\frac{\nu_m}{\nu_a^{(2)}}\right) \left(\frac{\nu}{4 \g^2 \nu_m x_0}\right)^{-\frac{p}{2}}, & \nu > 4 \g^2 \nu_m x_0.
\end{array} \right.
\end{equation}
The integration over different electron energies again needs to be
divided into four different regimes:
\begin{eqnarray}
\label{int2} f_{\nu}^{IC} = R \sigma_T \\
\nonumber &\times& \left\{\begin{array}{ll}

\displaystyle  \int_{\g_c}^{\infty}{d \g N (\g) I_1}+, & \nu < \nu_a^{IC}; \\

\left[\displaystyle \int_{\g_c}^{\g_{cr} (\nu_a)}{d \g N (\g) I_2} + \displaystyle \int_{\g_{cr} (\nu_a)}^{\infty}{d \g
N (\g) I_1}\right], & \nu_a^{IC} < \nu < \nu_m^{IC}; \\

\left[\displaystyle \int_{\g_c}^{\g_{cr} (\nu_m)}{d \g N (\g) I_3} + \displaystyle \int_{\g_{cr} (\nu_m)}^{\g_{cr}
(\nu_a)}{d \g N (\g) I_2} + \displaystyle \int_{\g_{cr} (\nu_a)}^{\infty}{d \g N (\g) I_1}\right], &
\nu_m^{IC} < \nu ; \\

\end{array} \right.
\end{eqnarray}

Evaluating the integrals in Eqn. (\ref{int2}), we only keep the
dominant terms:
\begin{eqnarray}
\label{fc2} f_{\nu}^{IC} &\simeq& R \sigma_T n f_{max} x_0 \\
\nonumber &\times&\left\{ \begin{array}{ll}

\frac{5}{9} \left(\frac{\nu}{\nu_a^{IC}}\right), & \nu < \nu_a^{IC} \\

\frac{1}{3} (\frac{\nu}{\nu_{a}^{IC}})^{-1/2}[\log\left(\frac{1}{3}\frac{\nu}{\nu_a^{IC}}\right)+\frac{5}{9}], &
\nu_a^{IC} < \nu < \sqrt{\nu_m^{IC} \nu_c^{IC}}; \\

(\frac{\nu}{\nu_a^{IC}})^{-1/2} [\frac{2}{(p+2)(p-1)} + \frac{1}{3}\log\left(
 \frac{\nu_m}{\nu_a}\right)+\frac{5}{9} ], &\sqrt{\nu_m^{IC}
 \nu_c^{IC}} < \nu < 4\g_m^2 \nu_a x_0; \\

(\frac{\nu}{\nu_a^{IC}})^{-1/2}[\frac{2}{(p+2)(p-1)} + \frac{2}{3(p-1)}+
\frac{1}{3}\log\left(\frac{\nu_m^{IC}}{\nu}\right)], &4\g_m^2
\nu_a x_0 < \nu < \nu_m^{IC}; \\

(\frac{\nu_m^{IC}}{\nu_a^{IC}})^{-1/2}(\frac{\nu}{\nu_m^{IC}})^{-p/2}[\frac{2}{(p+2)(p-1)}+
\frac{1}{p+2}\log\left(\frac{\nu}{\nu_m^{IC}}\right) +\frac{2}{3(p-1)}],  &4\g_m^2 \nu_a x_0 < \nu < \nu_m^{IC}
\end{array} \right.
\end{eqnarray}

\subsection{$\nu_c < \nu_m < \nu_a$}

\begin{equation}
\label{int12} I = \left\{ \begin{array}{l@{\quad \quad}l}
I_1\simeq\frac{2(p+5)}{3(p+2)} f_{max} x_0 \left(\frac{\nu}{4 \g^2 \nu_a x_0}\right), &  \nu < 4 \g^2 \nu_a^{(2)} x_0 \\
I_2 \simeq \frac{2}{p+2} f_{max} x_0 \left(\frac{\nu}{4 \g^2 \nu_m x_0}\right)^{\frac{-p}{2}},& x_0 < \nu > 4 \g^2 \nu_a x_0 \\
\end{array} \right.
\end{equation}

\begin{eqnarray}
\label{fc22} f_{\nu}^{IC} &\simeq& R \sigma_T n f_{max} x_0 \\
\nonumber &\times&\left\{ \begin{array}{ll}

\frac{2(p+5)}{9(p+2)} \left(\frac{\nu}{\nu_a^{IC}}\right), & \nu < \nu_a^{IC} \\

(\frac{\nu}{\nu_{a}^{IC}})^{-1/2} [\frac{2}{(p+2)(p-1)}+ \frac{2(p+5)}{9(p+2)}],
  & \nu_a^{IC} < \nu < 4 \g_m^2 \nu_a x_0; \\

 (\frac{\nu}{\nu_a^{IC}})^{-p/2}( \frac{\g_m}{\g_c})^{(p-1)}
[\frac{2}{(p+2)(p-1)}+\frac{1}{(p+2)}\log\left(\frac{\nu}{4 \g_m^2 \nu_a x_0}\right)+\frac{2(p+5)}{3(p+2)^2}],
 & \nu_c^{IC} < \nu ;
\end{array} \right.
\end{eqnarray}

\subsection{$\nu_m < \nu_c < \nu_a$}

\begin{equation}
\label{int13} I = \left\{ \begin{array}{l@{\quad \quad}l}
I_1\simeq\frac{2(p+5)}{3(p+2)} f_{max} x_0 \left(\frac{\nu}{4 \g^2 \nu_a x_0}\right), &  \nu < 4 \g^2 \nu_a x_0 \\
I_2 \simeq \frac{2}{p+2} f_{max} x_0 \left(\frac{\nu}{4 \g^2 \nu_m x_0}\right)^{\frac{-p}{2}},& \nu > 4 \g^2 \nu_a x_0 \\
\end{array} \right.
\end{equation}

\begin{eqnarray}
\label{fc23} f_{\nu}^{IC} &\simeq& R \sigma_T n f_{max} x_0 \\
\nonumber &\times&\left\{ \begin{array}{ll}

\frac{2(p+5)(p-1)}{3(p+2)(p+1)} \left(\frac{\nu}{\nu_a^{IC}}\right), & \nu < \nu_a^{IC} \\

 (\frac{\nu}{\nu_{a}^{IC}})^{-{\frac{(p-1)}{2}}}[\frac{2(p-1)}{(p+2)}+\frac{2(p+5)(p-1)}{3(p+2)(p+1)}],
&  \nu_a^{IC} < \nu < 4 \g_c^2 \nu_a x_0 \\

 (\frac{\nu}{\nu_{a}^{IC}})^{-p/2}(\frac{\g_c}{\g_m}) [\frac{2(p-1)}{(p+2)}+ \frac{(p-1)}{(p+2)}\log\left(\frac{\nu}{4 \g_c^2\nu_a x_0}\right) +
 \frac{2(p+5)(p-1)}{3(p+2)^2}],
&  \nu > 4\g_c^2 \nu_a x_0;
\end{array} \right.
\end{eqnarray}

\subsection{$\nu_m < \nu_a < \nu_c$}

\begin{equation}
\label{int14} I = \left\{ \begin{array}{l@{\quad \quad}l}
I_1\simeq\frac{2(p+4)}{3(p+1)} f_{max} x_0 \left(\frac{\nu}{4 \g^2 \nu_a^{(2)} x_0}\right), &  \nu < 4 \g^2 \nu_a^{(2)} x_0 \\
I_2 \simeq \frac{2}{(p+1)} f_{max} x_0 \left(\frac{\nu}{4 \g^2 \nu_m x_0}\right)^{\frac{-(p-1)}{2}},& 4 \g^2 \nu_a^{(2)} x_0 < \nu < 4 \g^2 \nu_m x_0 \\
I_3 \simeq \frac{2}{(p+2)} f_{max} x_0 \left(\frac{\nu_c}{\nu_a}\right)^{-(p-1)/2} \left(\frac{\nu}{4 \g^2 \nu_c
x_0}\right)^{-\frac{p}{2}}, & \nu > 4 \g^2 \nu_m x_0.
\end{array} \right.
\end{equation}

\begin{eqnarray}
\label{fc24} f_{\nu}^{IC} &\simeq& R \sigma_T n f_{max} x_0 \\
\nonumber &\times&\left\{ \begin{array}{ll}

\frac{2(p-1)(p+4)}{3(p+1)^2} \left(\frac{\nu}{\nu_a^{IC}}\right), & \nu < \nu_a^{IC} \\

(\frac{\nu}{\nu_{a}^{IC}})^{-(p-1)/2}[\frac{(p-1)}{(p+1)}
\log\left(\frac{\nu}{\nu_a^{IC}}\right)+\frac{2(p+4)(p-1)}{3(p+1)^2} ],
&  \nu_a^{IC} < \nu < \sqrt{\nu_m^{IC} \nu_c^{IC}}; \\

 (\frac{\nu}{\nu_a^{IC}})^{-(p-1)/2} [\frac{2(p-1)}{(p+2)}+ \frac{(p-1)}{(p+1)}\log\left(\frac{\nu_c}{\nu_a}\right)+
 \frac{2(p+4)(p-1)}{3(p+1)^2}],
 &\sqrt{\nu_m^{IC} \nu_c^{IC}} < \nu < 4\g_c^2 \nu_a x_0; \\

 (\frac{\nu}{\nu_a^{IC}})^{-(p-1)/2}[\frac{2(p-1)}{(p+2)}+ \frac{(p-1)}{(p+1)} \log\left(\frac{\nu_c^{IC}}{\nu}\right)+
 \frac{2(p-1)}{(p+1)}], &4\g_c^2 \nu_a x_0 < \nu < \nu_c^{IC}; \\

(\frac{\nu_c^{IC}}{\nu_a^{IC}})^{-(p-1)/2}(\frac{\nu}{\nu_c^{IC}})^{-p/2}[\frac{2(p-1)}{(p+2)}+\frac{(p-1)}{p+2}\log\left(\frac{\nu}{\nu_c^{IC}}\right)+\frac{2(p-1)}{(p+1)}],&\nu
> \nu_c^{IC}
\end{array} \right.
\end{eqnarray}

It is noted that a factor of $1/(p-1)$ has been multiplied for the
slow-cooling case in order to keep the IC flux continuous while the
afterglow changes from the fast-cooling regime to the slow-cooling
regime.


\section{Derivation of Radiative Correction Factor}
\label{app:der_rad}
In the observer frame the energy loss rate is equal to the rate at
which energy is supplied to the unshocked matter $4\pi R^2 p_2$
\citep{cps+98}, multiplied by $\epsilon$, the fraction of energy that
each particle has lost (assuming that the relation between the radius
and the Lorentz factor is described by $R=A\Gamma^2 ct$ where A=2 for
the GRB prompt phase and $A \in [3,7]$ for the afterglow deceleration
phase).

\begin{equation}
\frac{dE}{dt}=-4 \pi R^2 p_2 \epsilon = -4 \pi A^2 \Gamma^4 c^2 t^2 p_2
\epsilon
\label{eqn:energy_loss}
\end{equation}
where $R$ is the radius of the shocked region from the center in the
observer frame, $p_2=c \Gamma^2 U'/3$ is the pressure in the shocked
region in the observer frame, and $U'=2 \Gamma^2 n m_p c^2$ is the
energy density of the shocked region in the comoving frame.

Normally we assume that all the energy stored in electrons has been
radiated during the fast cooling phase, i.e.,
$\epsilon=\epsilon_e$. However, \citet{cps+98} have shown that only a
portion of the energy will be lost even if the electrons are in the
fast-cooling regime (refer to their Fig.~6). The relation between the
radiation factor $\epsilon$ and the electron equipartition parameter
$\epsilon_e$ is given by
$\epsilon_e=1-(\frac{1-\epsilon}{\sqrt{1+\epsilon}})^{\hat{\gamma}}$
(their Eq.~46) where $\hat{\gamma}=4/3$ for the extreme relativistic
case is the adiabatic index. We can solve the above equation
numerically, e.g., $\epsilon=0.11$ for $\epsilon_e=0.2$,
$\epsilon=0.0255$ for $\epsilon_e=0.05$, and $\epsilon=0.005$ for
$\epsilon_e=0.01$. We find that for the reasonable range of parameter
values, we have the following relation, $\epsilon \sim \frac{1}{2}
\epsilon_e$, and substituting it and BM solution back into
Eq.~\ref{eqn:energy_loss} above, we obtain:
\begin{eqnarray}
\frac{dE}{dt}&=& -\frac{17}{12 A} \epsilon_e \frac{E}{t}
\label{eqn:lum_rate}
\end{eqnarray}

We have the solution as $E(t)=E_0 (\frac{t}{t_0})^{-17\epsilon_e/12A}$
where $t_0$ is the deceleration time. Because we have applied the
Blandford-McKee solution during the whole derivation, the above
solution is only valid after the deceleration time. Normally during
the afterglow deceleration phase, the we set the parameter $A=4$, and
thus we obtain the radiative correction factor after prompt phase
$R=(t/t_0)^{17\epsilon_e /48}$.  We note that the above correction
factor can only be applied to the fast cooling case. Taking
$\epsilon_e=0.1$ and $t_0=300$ s, the correction factor at observer
time $t=10$ hours would be $R\sim 1.18$. It should be mentioned that
if the estimated correction factor is much larger than the unity, one
should not estimate the kinetic energy only by introducing the
radiative correction factor $R$ for the adiabatic solution, and should
reconsider the whole evolution including the radiation loss
effects. Because at this point the radiation is not negligible any
more, the afterglow evolution will deviate substantially from the BM
solution \citep[e.g., ][]{yhsf+03, wdhl+05}.

The radiative correction factor $R=(t/t_0)^{17\epsilon_e /48}$ for the
afterglow is different from the one, $R=(t/t_0)^{17\epsilon_e /16}$,
provided by \citet{s+97} (see also \citealt{lz04}) for two reasons:
(1) They have taken the value $A=16$, while we have used $A=4$
which is believed to be much more reasonable from the detailed studies
of the hydrodynamic evolution of the afterglow
\citep{wax+97,pm+97}. (2) They have assumed all the electron energy
will be radiated away if the afterglow is in the fast-cooling regime,
but from the results of \citet{cps+98}, they showed that roughly half
of the electron energy is lost.


\section{Kinetic Energy}

If the \xray\ band is above the typical frequency $\nu_m$ and the
cooling frequency $\nu_c$, the observed flux at a certain observer
time is given by:
\begin{eqnarray}
\nonumber
F_{\nu,X}&=&F_{\nu,m} \nu_m^{(p-1)/2} \nu_c^{1/2} \nu^{-p/2} \\
\nonumber &=& 2^{(2-3p)/4} \times 10^{-30}\ {\rm erg s^{-1} cm^{-2}
Hz^{-1}}
\\ && \times (1+z)^{(p+2)/4} E_{52}^{(p+2)/4} \epsilon_{e,-1}^{(p-1)}
\epsilon_{B,-2}^{(p-2)/4} t_{10 {\rm h}}^{(2-3p)/4} \nu_{18}^{-p/2} (1+Y)^{-1}
D_{L,28}^{-2}
\end{eqnarray}
where $p$ is the energy distribution index, $E_{52}$ is the
isotropic-equivalent kinetic energy, $\epsilon_e$ is the electron
equipartition parameter, $\epsilon_{B}$ is the magnetic field
equipartition parameter, $t$ is the observer time, $\nu$ is the X-ray
observing frequency, and $Y$ is the Compton parameter.

The \xray\ luminosity  is described as \citep{lr+00,bkf03}:
\begin{equation}
L_{\nu,X}=4 \pi D_{L}^2 F_{\nu,X} (1+z)^{-\alpha+\beta-1} \\
\end{equation}
and
\begin{eqnarray}
\nonumber L_{X}&=& \int L_{\nu,X} d\nu \\ \nonumber &=& C \nu L_{\nu,X} \\
&\approx& 2.5 \times 10^{45}~{\rm erg~s^{-1}}\ (1+z)^{(p/4-1/2-\alpha
+\beta)} \nonumber \\& & \times E_{52}^{(p+2)/4} \epsilon_{e,-1}^{(p-1)}
\epsilon_{B,-2}^{(p-2)/4} t_{10 {\rm h}}^{(2-3p)/4} \nu_{18}^{(2-p)/2}
(1+Y)^{-1}
\end{eqnarray}
where $C \equiv\int L_{\nu,X} d\nu/\nu L_{\nu,X} $ is an integration
constant, and it shows that $C \approx 4.3$.

We can find the kinetic energy $E$ from the above equation reversely,
and considering the radiative correction factor, we obtain,
\begin{eqnarray}
\nonumber E_{K} &=& 10^{52} {\rm ergs} \  R [\frac{L_{X}}{2.5 \times
10^{45} {\rm ergs \ s^{-1}}}]^{4/(p+2)} (1+z)^{4(p/4-1/2-\alpha
+\beta)/(p+2)} \\ && \times \epsilon_{e,-1}^{4(p-1)/(p+2)}
\epsilon_{B,-2}^{-(p-2)/(p+2)} t_{10 {\rm h}}^{(2-3p)/(p+2)}
\nu_{18}^{2(2-p)/(p+2)} (1+Y)^{4/(p+2)} \label{eqn:kinetic_energy}
\end{eqnarray}
where $R=[\frac{t}{t_{dec}}]^{(17/48) \epsilon_e}$ is the radiative
correction factor, and the derivation is given in Appendix
\ref{app:der_rad}.

\end{document}